\documentclass[lettersize,journal]{IEEEtran}

\usepackage{booktabs}
\usepackage{makecell}

\renewcommand{\arraystretch}{1.3}

\usepackage[numbers]{natbib}
\usepackage{amsmath,amssymb,amsfonts}
\usepackage{hyperref}
\usepackage{algorithmic}
\usepackage{float}
\usepackage{graphicx}
\usepackage{textcomp}
\usepackage{xcolor}
\usepackage{multirow}
\usepackage{graphicx}
\usepackage{adjustbox}
\usepackage{longtable}
\usepackage[normalem]{ulem}
\usepackage{caption}
\usepackage{tabularx}
\usepackage{listings}

\begin{document}

\title{A Unified Feature Model for Microservice Identification and Refactoring}

\author{\IEEEauthorblockN{Ana Almeida and António Rito Silva} \\
\IEEEauthorblockA{\textit{INESC-ID, Instituto Superior Técnico} \\
\textit{University of Lisbon}\\
\{ana.margarida.almeida@tecnico.ulisboa.pt, rito.silva@tecnico.ulisboa.pt\}}
}


\maketitle

\begin{abstract}
Several approaches have been proposed for the automatic identification of microservices within monolithic systems. These methodologies differ in their data collection and analysis techniques, the decomposition algorithms applied, and the mechanisms used to visualize and refine candidate microservices. Despite this diversity, systematic experimentation and comparison across approaches remain limited by the absence of a common conceptual framework. Furthermore, current research indicates that no single optimal method exists; rather, integrating multiple approaches is necessary to fully explore the trade-offs inherent in any design solution.

To address this gap, this paper proposes a feature model for variant-rich microservice identification tools, grounded in an extensive analysis of the state of the art. We evaluate this feature model through a systematic mapping of representative literature and by analyzing its instantiation within the architecture of an existing microservice identification tool. Our findings are two-fold. First, the proposed feature model successfully captures the primary variation points of existing methodologies, providing a unifying foundation for analyzing, comparing, and designing microservice identification tools. Second, while no individual tool covers more than a fraction of the model, the analyzed tools jointly span almost all of its variation points. This complementarity is the central result: the design space is already populated, but it is fragmented across standalone tools, none of which can compare or integrate the alternatives that the others implement.
\end{abstract}


\section{Introduction}

At the early stages of application development, many teams adopt a monolithic architecture due to its simplicity of implementation. Monolithic systems benefit from a shared domain model managed by a single transactional system, which simplifies the development of application business logic~\cite{InDefenseOfTheMonolith17}.
However, as systems grow in size and complexity, monolithic architectures increasingly hinder independent scalability and do not support development by small, autonomous teams. The tight coupling imposed by a shared codebase and data model slows down development and deployment, limits the adoption of agile practices, and complicates the migration to new technologies. Over time, these limitations cause monolithic systems to become harder to maintain and evolve~\cite{Werner2006}.

Microservice architectures, on the other hand, allow the decomposition of systems into smaller, independently deployable services, each aligned with a specific business capability. This architectural style improves scalability, maintainability, and organizational flexibility by enabling teams to develop, deploy, and evolve services independently~\cite{FowlerMicroservices, Thones15}.

Driven by these benefits, microservice architectures have emerged as a widely adopted alternative~\cite{Thones15} and an increasing number of monolithic systems are being migrated to microservices~\cite{ChrisRichardson17}.

However, this migration is a complex and challenging task. One of the most difficult problems is the identification of the microservices that should result from monolith decomposition~\cite{Abdellatif2021}. As monolithic systems grow larger, the identification problem becomes harder due to the size of the codebase, the complexity of dependencies, and the need for substantial domain knowledge.

To address this challenge, several approaches have been proposed for identifying microservices within monolithic systems, e.g.~\cite{gysel2016service, jin21}. These approaches differ in terms of the data they collect from the monolith, how they analyze it, and the techniques they apply. For instance, some rely on static analysis of the source code~\cite{selmadji2018re,Nunes19}, others analyze runtime behavior through dynamic analysis~\cite{fuhr2011using,jin21}, while others operate on higher-level models of the system~\cite{gysel2016service,tyszberowicz2018identifying}. Similarly, decomposition strategies vary widely, with some approaches clustering classes or persistent domain entities, e.g.~\cite{selmadji2018re, tyszberowicz2018identifying, Nunes19}, and others focusing on business functions, e.g.~\cite{zhang2005service, saha2015service}. As a result, different techniques prioritize different quality attributes and often produce distinct decompositions for the same system.

Despite the proliferation of proposed techniques, existing approaches are typically implemented as standalone solutions that support only a limited set of analysis methods, decomposition algorithms, and evaluation metrics. This fragmentation presents two main drawbacks. First, comparative analysis is highly impractical because these tools differ significantly in everything from data collection to assessment metrics; researchers struggle to systematically evaluate them against one another. Second, an effective solution requires a trade-off analysis of the diverse, and often conflicting, quality attributes desired in the final microservice architecture, such as performance and modularity, necessitating the integration of multiple approaches.

Additionally, the sheer volume of secondary studies e.g.~\cite{Francesco2018, Fritzsch2018, Ponce2019, Abdellatif2021, Auer2021, Capuano2022, Abgaz2023, Habib24, Oumoussa2024, Wang2024, Mehmood25, Mohottige2025, Saucedo2025} in under a decade underscores a field that is expanding rapidly but lacks a unified direction.

The aforementioned surveys and systematic literature reviews are primarily descriptive and require significant preprocessing to normalize results across techniques. Additionally, they do not provide a tool to support experimentation, comparison, or even the integration of approaches. Moreover, they lack a common analysis framework, which, in some sense, is reflected by the large number of secondary studies in such a short period.

Taken together, the large number of diverse and heterogeneous approaches proposed for microservice identification highlights that the main challenge in the contemporary landscape is no longer the lack of decomposition strategies but rather the absence of a common analysis framework that supports systematic experimentation, comparison, and integration of these strategies.

Our work addresses the aforementioned problem by proposing a feature model~\cite{Zhang2004}, a technique typically used in Software Product Lines~\cite{Benavides2010, Kastner2013, Apel2013}, that captures the variability of microservice identification and refactoring approaches because the identification process is inherently variable. Therefore, we argue that it is necessary to shift the analysis from the solution space to the problem space to establish a unified analysis framework for identifying microservices within monolithic systems.

The feature model design results from an extensive analysis of the state of the art and is intended to serve as a requirements foundation for the design of variant-rich microservice identification tools. To do so, it has to represent the mandatory and optional features necessary to support the complete microservice identification process, including data extraction, analysis, decomposition, visualization, editing, and comparison.

The process of modeling this feature model follows the principles proposed by Nešić et al.~\cite{Nesic2019}, which structured our methodology into different phases: a search phase to identify and select the relevant literature; a top-down modeling phase, where domain experts define more general and abstract features based on a known pipeline; a bottom-up modeling phase to refine and extend the model by analyzing diverse approaches; and an evaluation phase to assess the model's completeness and applicability. This methodology is further described in Section~\ref{sec:modeling-process}.

The feature model benefits the research community by explicitly defining the design space of microservice identification, highlighting open variation points, and enabling systematic experimentation and comparison of future decomposition strategies.

By making variability explicit, the feature model also supports the software development community by guiding the design and extension of microservice identification tools, as well as assisting software architects in exploring and evaluating alternative decompositions according to system, domain, and migration requirements.

Therefore, our contributions are:

\begin{itemize}
    \item \textit{A Unified Feature Model}: Derived from an extensive meta-analysis of the state of the art, this model represents the essential variation points of the identification process, including extraction, analysis, decomposition, and visualization.
    \item \textit{Requirements Foundation}: We provide a blueprint for the design of variant-rich identification tools, allowing developers to build modular systems that can swap algorithms or metrics as needed.
    \item \textit{A Framework for Multi-Objective Trade-off Analysis}: We explicitly model the conflicting quality attributes (e.g. performance vs. modularity) inherent in microservice design. This allows architects to integrate multiple, disparate approaches, such as combining static dependency analysis with business-process mining, to find decompositions that balance competing requirements.
    \item \textit{A Roadmap for Benchmarking}: By making variability explicit, our model enables the research community to define a common design space, highlighting unexplored areas and facilitating the systematic comparison of future strategies.
\end{itemize}


The remainder of this report is organized as follows. 
Section~\ref{sec:background} introduces the background concepts required to frame this work. It discusses monolithic and microservices architectures, the monolith-to-microservices migration process, and the microservices identification problem (Section~\ref{sec:monolith-to-microservices-migration}). It also presents the \textit{Mono2Micro} tool (Section~\ref{sec:mono2micro}) and introduces feature models as the modeling technique adopted in this work (Section~\ref{sec:feature-model-background}). 
Section~\ref{sec:research-questions} presents the research questions and the methodology followed to construct the proposed feature model. It describes the modeling process, the principles that guided the feature model design, and the systematic mapping process used to identify, analyze, and consolidate the variability found in the literature.
Section~\ref{sec:feature-model-structure} presents the resulting feature model. It introduces the model incrementally, describing its main feature groups and variation points.
Section~\ref{sec:evaluation} evaluates the proposed feature model, trying to analyze its strengths and limitations by mapping representative microservice identification tools onto the model.
Section~\ref{sec:related-work} discusses the related work, comparing the proposed feature model with existing mapping studies and systematic literature reviews on microservice identification and migration.
Section~\ref{sec:threats-to-validity} discusses the threats to validity associated with the construction and evaluation of the feature model, including construct, internal, external, and conclusion validity.
Finally, Section~\ref{sec:conclusion} concludes the report by summarizing the main contributions of this work and discussing directions for future research.

\section{Background}
\label{sec:background}

This section introduces three background concepts required to frame the proposed solution:

\begin{itemize}
    \item \textbf{Monolith-to-microservices migration}, which contextualizes the problem addressed in this work.
    \item \textbf{Mono2Micro}, an extensible tool that addresses the same problem space and is used in this work as a reference and evaluation platform.
    \item \textbf{Feature models}, the modeling technique adopted to represent the proposed solution.
\end{itemize}

\subsection{Monolith-to-Microservices Migration}
\label{sec:monolith-to-microservices-migration}

\subsubsection{Monolithic Architecture}

A monolithic architecture is a software architectural style in which all components are developed, deployed, and executed as a single unit. In these systems, a shared codebase and a centralized data model support the business logic and user interface, while communication between components typically occurs through in-process method calls. This design provides good performance and simplified deployment, as the application is delivered as a single executable~\cite{FowlerMicroservices}.

Due to these characteristics, monolithic architectures are commonly adopted in the early stages of software development, since the shared domain model simplifies initial design decisions and facilitates consistency across the system, ensuring rapid development and straightforward evolution during the initial phases~\cite{InDefenseOfTheMonolith17}.

However, as systems grow in size and complexity, the tightly coupled structure of monolithic applications does not support the development by small, autonomous teams; a change to a single component may require organizational approval and imply rebuilding and redeploying the entire system. Furthermore, monolithic architectures do not support independent scalability, as all components must be scaled together, even when only specific functionalities require additional resources.

These limitations negatively impact maintainability by reducing development agility, extending release cycles, and hindering the adoption of new technologies and architectural practices.

\subsubsection{Microservices Architecture}

A microservices architecture is a software architectural style in which an application is decomposed into multiple small, loosely coupled, and independently scalable services. Each microservice is responsible for a specific business capability, manages its own data, and communicates with other services through lightweight communication mechanisms, such as RESTful APIs or messaging systems~\cite{Thones15}.

A key principle of microservice architecture is that, unlike monolithic applications, it is possible to deploy each service independently. This allows each business capability to be developed at its own team's pace with controlled impact on teams responsible for other business capabilities. This enables agile software development and facilitates the adoption of new technologies or architectural practices. Additionally, it improves scalability, as each service can be replicated and scaled independently based on its specific workload requirements, allowing for more efficient resource utilization by allocating computational resources only where they are needed.

Despite these advantages, microservice architectures introduce additional complexity and may lead to performance overhead due to the distributed communication and require careful handling of data consistency and fault tolerance across services~\cite{richardson2018}. Moreover, designing and maintaining a microservice-based system demands greater architectural effort to ensure that service boundaries are well-defined and that the overall system remains coherent.

Consequently, the benefits and drawbacks of adopting a microservice architecture instead of a monolithic architecture must be carefully evaluated, as they strongly depend on system requirements and the quality of the service decomposition.

\subsubsection{Migration}

In many real-world scenarios, existing monolithic systems need to be transformed into microservice architectures~\cite{Francesco2018, Abdellatif2021, Abgaz2023}. However, this migration process is inherently complex, as it often involves large and long-lived systems characterized by tightly coupled components, complex dependencies, and limited or outdated documentation. These factors significantly hinder system comprehension and increase the need for domain expertise during the migration process~\cite{Abdellatif2021, Francesco2018, Capuano2022}.

One of the main challenges in monolith-to-microservices migration is the identification of appropriate microservices within the monolithic system~\cite{Abdellatif2021, Abgaz2023, Mohottige2025, Saucedo2025}. Poorly defined service boundaries can undermine the expected benefits of microservice architectures and increase their drawbacks.

Another major challenge concerns the migration process itself. In practice, migrations must often be performed while the system remains operational and continues to evolve. To address this challenge, incremental migration strategies have been proposed, among which the \textit{Strangler Fig Pattern} is one of the most widely adopted~\cite{FowlerStrangler}. This pattern fosters the gradual replacement of parts of the monolithic system with microservices while the original system remains in operation. This approach reduces migration risks and enables continuous system evolution, as new functionality is implemented as microservices and existing functionality is progressively extracted.

Nevertheless, the successful application of such strategies requires prior knowledge about which parts of the system should be extracted and how they should be grouped into microservices. Consequently, the problem of microservice identification becomes a central concern in monolith-to-microservices migration, directly influencing both the technical and organizational outcomes of the transformation.

\subsubsection{Microservices Identification}

Microservice identification is the process of defining suitable service boundaries within an existing system. Although it is most commonly applied to monolithic applications, it can also be used to refine and restructure existing microservice architectures~\cite{Zhong2025}.

Independent of the specific techniques employed, and as extensively reported in several studies on the identification of microservices in monolithic systems, e.g.~\cite{Abdellatif2021, Abgaz2023, Oumoussa2024}, the process involves addressing a set of recurring concerns. First, relevant information about the system must be collected. This information may originate from different sources, including source code, e.g.~\cite{Santos22}, database schemas, e.g.~\cite{Romani22}, execution traces, e.g.~\cite{Jin18}, development history, e.g.~\cite{Lourenco23}, or higher-level models such as business process descriptions, e.g.~\cite{Taibi19}.

Second, the collected data must be analyzed to extract relevant information about the system that can support decomposition decisions. Depending on the available data, different analysis techniques may be applied, resulting in different representations, such as dependency graphs, e.g.~\cite{Desai21}, call graphs, e.g.~\cite{Nunes19}, or usage frequency metrics, e.g.~\cite{Bajaj20}.

Third, decomposition decisions are made based on the analysis results. These decisions may rely on different types of information, employ distinct algorithms or optimization techniques, and target different quality attributes, such as maintainability, e.g.~\cite{Barbosa20}, scalability, e.g.~\cite{Ahmadvand16}, or performance, e.g.~\cite{Mustafa18}. As a result, different approaches may produce alternative decompositions for the same system, each with distinct trade-offs.

Finally, to assess the quality of the generated decompositions, evaluation mechanisms are employed. These typically include quantitative metrics, such as service size, coupling, cohesion, complexity, and other measures that can be computed based on the initially collected data~\cite{Bogner2017}. Comparative analyses across multiple decompositions or benchmarking techniques can also be used to gain further insights, e.g.~\cite{Andrade23}.

In addition, visualization and editing mechanisms might also be important to support architects in inspecting candidate decompositions, incorporating domain knowledge, and iteratively refining service boundaries, e.g.~\cite{Nakazawa2018}.

Overall, it is possible to observe an inherent variability of microservice identification approaches across all these concerns, including data sources, analysis techniques, decomposition strategies, evaluation criteria, and even the use of other additional supporting mechanisms such as visualization and editing.

\subsection{Mono2Micro Tool}
\label{sec:mono2micro}

Mono2Micro~\cite{Lopes2023} is an extensible multiple-strategy tool designed to support the comparison of decomposition approaches and help future investigators in the quality assessment of new decomposition approaches~\cite{Lopes2023}. The tool focuses only on monolith-to-microservices migration; it does not address the refactoring of existing microservices architectures.

\begin{figure*}
    \centering
    \includegraphics[width=1\linewidth]{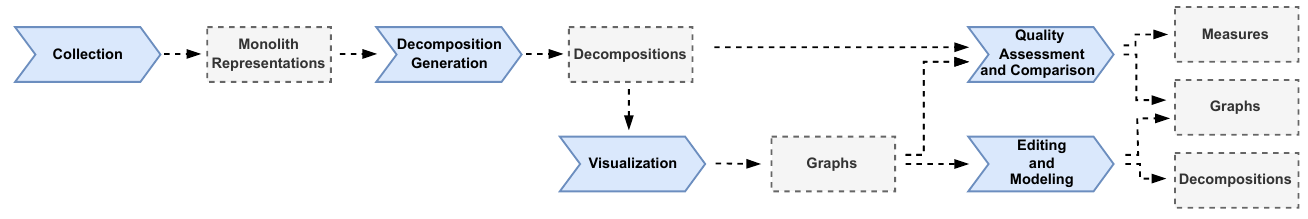}
    \caption{Microservices identification pipeline as proposed in~\cite{Lopes2023}}
    \label{fig:pipeline}
\end{figure*}

Based on an analysis of existing microservice identification approaches, Lopes and Silva proposed a microservice identification pipeline (Figure~\ref{fig:pipeline}) that consolidates the common stages observed in the literature, including data collection, analysis, decomposition, evaluation, visualization, and editing. \textit{Mono2Micro} implements this pipeline and supports a limited number of identification strategies.

A key design principle of \textit{Mono2Micro} is extensibility. The tool was designed to allow the integration of new strategies that may rely on different data sources, analysis techniques, decomposition algorithms, and evaluation mechanisms. This makes \textit{Mono2Micro} a relevant platform for experimenting with microservice identification approaches and comparing them within a unified environment.

These characteristics make \textit{Mono2Micro} a representative example of a microservice identification tool that explicitly addresses variability across the identification pipeline, aligning with our research goals. However, despite these strengths, \textit{Mono2Micro} currently supports a limited set of strategies and design choices. Several aspects explored in the literature, such as the use of additional data sources~\cite{Wang2024}, alternative evaluation criteria~\cite{Wang2024}, dynamic adaptation of service granularity~\cite{Hassan2022}, or microservice-to-microservice refactoring scenarios~\cite{Zhong2025}, are not fully supported.

\subsection{Feature Model}
\label{sec:feature-model-background}

Feature models are a widely used modeling technique originating from the Software Product Line (SPL) engineering paradigm~\cite{Benavides2010, Kastner2013, Apel2013}. Their goal is to represent the commonalities and variabilities of a family of related systems by capturing the set of features that characterize a system domain and the relationships between them.

In general, a feature model represents a system in terms of features, where each feature represents a characteristic of the software system~\cite{Benavides2010}. Features are organized in a tree-like hierarchy, where the root represents the system itself, and higher-level features describe more abstract concepts, while lower-level features correspond to more concrete functionalities~\cite{Kastner2013}. It is possible to represent mandatory and optional features, as well as alternative (XOR) and inclusive (OR) feature groups~\cite{Benavides2010, Acher2010}. Additionally, to capture dependencies and incompatibilities between features that cannot be expressed through the hierarchy, cross-tree constraints, such as requires and excludes, can be used~\cite{Zhang2004}.

These characteristics make feature models particularly suitable for modeling systems with a high degree of variability, as demonstrated by Díaz et al.~\cite{Diaz2015}. In practice, feature models are used to analyze variability, reason about design alternatives, structure design spaces, and support the evolution of such systems.

\begin{figure}
    \centering
    \includegraphics[width=1\linewidth]{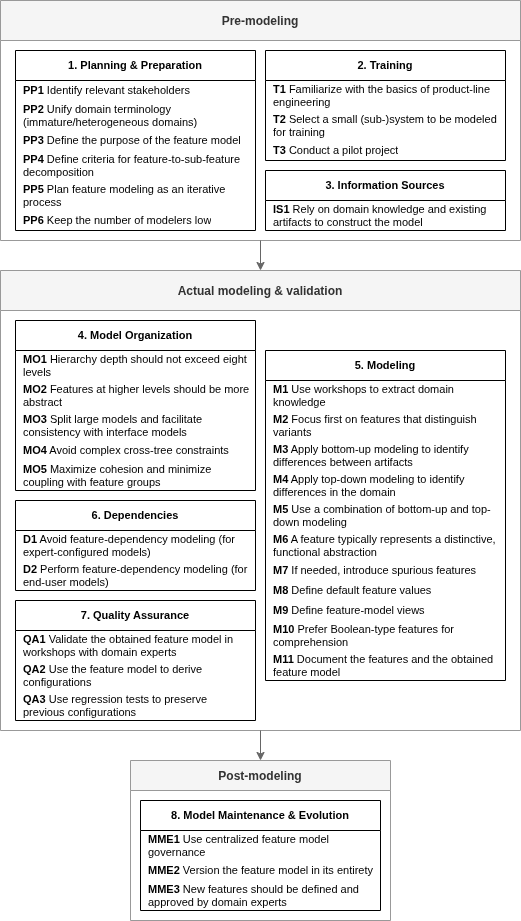}
    \caption{Principles for feature model construction~\cite{Nesic2019}}
    \label{fig:Nesic2019-principles}
\end{figure}

Building feature models is a complex engineering task that may involve multiple stakeholders, including domain experts, software architects, and tool developers. To support this process, Nešić et al.~\cite{Nesic2019} identified a set of principles that guide the planning, construction, and evolution of feature models, organized across three lifecycle activities (Figure~\ref{fig:Nesic2019-principles}).

In the \textbf{pre-modeling activities}, the \textit{Planning \& Preparation} (PP) principles require defining the model's purpose, identifying relevant stakeholders, unifying domain terminology, establishing decomposition criteria, and planning for iteration with a small number of modelers. The \textit{Training} (T) principles ensure modelers are familiar with product-line engineering before starting, while the single \textit{Information Sources} principle (IS) requires grounding the model in domain knowledge and existing artifacts.

During the \textbf{actual modeling and validation activities}, the \textit{Model Organization} (MO) principles govern the structure of the hierarchy, covering its depth, abstraction levels, cohesion, and use of cross-tree constraints. The \textit{Modeling} (M) principles cover the construction strategies, including top-down and bottom-up elicitation, feature type preferences, and documentation, while \textit{Dependencies} (D) principles address how feature relationships should be captured. \textit{Quality Assurance} (QA) principles require validating the model by deriving configurations and using regression references.

Finally, in the \textbf{post-modeling activities}, the \textit{Model Maintenance \& Evolution} (MME) principles govern versioning, centralized governance, and expert approval of new features. These principles are intended to improve the consistency and quality of the resulting model and have been validated across different application domains~\cite{Nesic2019}.

\section{Research Questions and Method}
\label{sec:research-questions}

The purpose of this work is to define a feature model that captures the commonalities and variability of monolith-to-microservice identification approaches. To achieve this goal, we followed a meta-review process, guided by the following research question:

\begin{itemize}
    \item RQ: How can we identify and represent the common parts and variability of the approaches for the identification of microservices in monolithic systems?
    \begin{itemize}
     \item What are the variability points?
     \item What are the common parts?
     \item How can these common and variable elements be represented in a feature model?
    \end{itemize}
\end{itemize}

\subsection{Modeling Process}
\label{sec:modeling-process}

\begin{figure}
    \centering
    \includegraphics[width=1\linewidth]{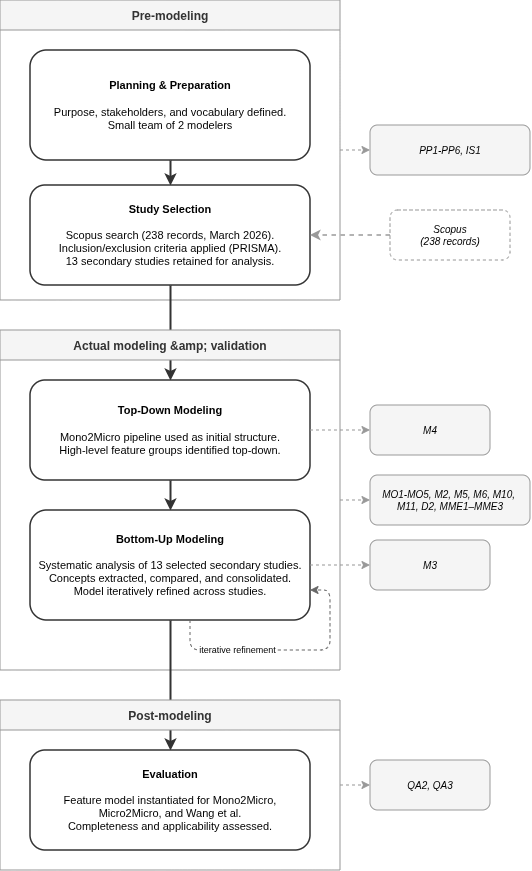}
    \caption{Methodology pipeline following Nešić et al.~\cite{Nesic2019}}
    \label{fig:methodology-pipeline}
\end{figure}

The feature modeling process adopted the principles of Nešić et al.~\cite{Nesic2019} (Figure~\ref{fig:Nesic2019-principles}) as described in Figure~\ref{fig:methodology-pipeline}.

\subsubsection{Planning \& Preparation}

The purpose of the model was explicitly defined before construction began (PP3): to provide a common analysis framework that supports systematic experimentation and comparison of microservice identification approaches. The relevant stakeholders, namely software developers and researchers interested in microservice identification, were identified (PP1), and a unified working vocabulary was established among the modelers to normalize the heterogeneous terminology encountered in the surveyed studies (PP2).

The process relied on domain knowledge and the analyzed artifacts as its primary information sources (IS1). It involved a small number of modelers (PP6), specifically a software engineering professor and a master's student. The process was planned as iterative from the outset (PP5), with centralized governance (MME1) maintained throughout: each version of the model was saved in its entirety using Git\footnote{\href{https://github.com/}{https://github.com/}} (MME2), and new features were discussed and approved by both modelers (MME3). Since both modelers are domain experts, training principles (T1-T3) and domain-knowledge-extraction workshops (M1) were not applicable.

\subsubsection{Study selection}

To identify publications that could support the elicitation of features relevant to monolith-to-microservice identification, we applied the inclusion and exclusion criteria presented in Table~\ref{tab:criteria}. Since the goal of this work is to derive a feature model rather than compare individual decomposition algorithms, the search focused on secondary studies, which are particularly suitable for identifying recurring concepts, dimensions, and classifications across the field.

\begin{table*}[htbp]
\caption{Study selection criteria}
\label{tab:criteria}
\centering
\begin{tabularx}{\textwidth}{@{}l X @{}} 
\toprule
\multicolumn{2}{l}{\textbf{Inclusion Criteria (IC)}} \\ \midrule
\textbf{ID} & \textbf{Criterion Description} \\ \midrule
IC1 & Study is a secondary study (e.g. SLR, survey, taxonomy, or mapping study). \\
IC2 & Study addresses monolith to microservice decomposition, microservice identification or closely related modernization approaches. \\
IC3 & Study provides sufficient methodological or conceptual detail to support feature extraction. \\\bottomrule
\multicolumn{2}{l}{\textbf{Exclusion Criteria (EC)}} \\ \midrule
\textbf{ID} & \textbf{Criterion Description} \\ \midrule
EC1 & Study is not written in English. \\
EC2 & Study is not peer-reviewed. \\
EC3 & Study is not published in a high-quality venue (CORE A* or A for conferences, Q1 or Q2 for journals). \\\bottomrule
\end{tabularx}
\end{table*}

A database search was conducted in Scopus in March 2026. The search string was designed to capture secondary studies related to monolith decomposition, microservice identification, and software modernization, resulting in the following query:

\begin{lstlisting}
("microservices" OR "microservice") AND ("monolith" OR "monolithic" OR "service identification" OR "migration" OR "refactoring" OR "reengineering" OR "modernization") AND ("survey" OR "taxonomy" OR "systematic literature review" OR "state-of-the-art" OR "mapping study")
\end{lstlisting}

This search returned 238 records. After applying the selection criteria, 11 studies were retained for analysis, as they were the ones that fully matched the selection criteria of this work.

\begin{figure}
    \centering
    \includegraphics[width=1\linewidth]{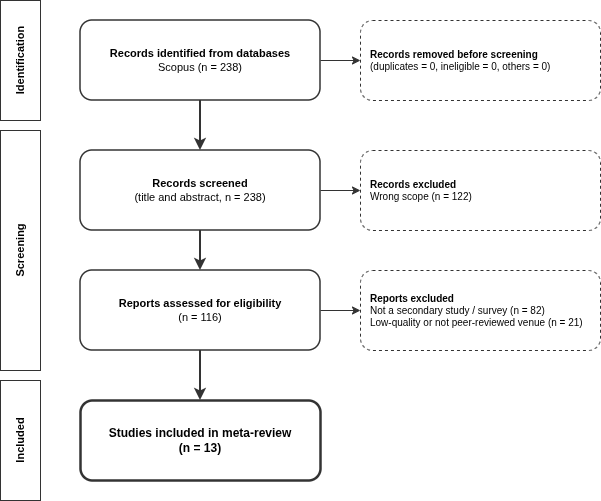}
    \caption{Selection process of the studies}
    \label{fig:selection-process}
\end{figure}

Figure~\ref{fig:selection-process} illustrates the selection process, and Table~\ref{tab:selected-studies} summarizes the 11 selected studies.

\begin{table*}[htbp]
\caption{Selected studies}
\label{tab:selected-studies}
\centering
\begin{tabularx}{\textwidth}{@{}l X l l l @{}}
\toprule
\textbf{Reference} & \textbf{Title} & \textbf{Study Type} & \textbf{Year} & \textbf{Ranking} \\ \midrule
\cite{Abdellatif2021} & A taxonomy of service identification approaches for legacy software systems modernization & Taxonomy & 2021 & Q1 \\
\cite{Abgaz2023} & Decomposition of Monolith Applications Into Microservices Architectures: A Systematic Review & SLR & 2023 & Q1 \\
\cite{Mohottige2025} & Reengineering software systems into microservices: State-of-the-art and future directions & SLR & 2025 & Q1 \\
\cite{Capuano2022} & A Systematic Literature Review on Migration to Microservices: a Quality Attributes perspective & SLR & 2022 & A \\
\cite{Saucedo2025} & Migration of monolithic systems to microservices: A systematic mapping study & Mapping Study & 2025 & Q1 \\
\cite{Oumoussa2024} & Evolution of Microservices Identification in Monolith Decomposition: A Systematic Review & SLR & 2024 & Q1 \\
\cite{Auer2021} & From monolithic systems to Microservices: An assessment framework & SLR & 2021 & Q1 \\
\cite{Nassima2025} & Semantic Approaches to Microservice Identification: A Systematic Literature Review & SLR & 2025 & Q1 \\
\cite{Oumoussa2025} & The Ontology-Based Mapping of Microservice Identification Approaches: A Systematic Study of Migration Strategies from Monolithic to Microservice Architectures & Mapping Study & 2025 & Q2 \\
\cite{Trabelsi2025} & A Systematic Literature Review of Machine Learning Approaches for Migrating Monolithic Systems to Microservices & SLR & 2025 & Q1 \\
\cite{Yang2025} & Full stack optimization of microservice architecture: systematic review and research opportunity & Systematic Review & 2025 & Q1 \\\bottomrule
\end{tabularx}
\end{table*}

\subsubsection{Top-Down Modeling}

In the top-down phase, we began by identifying the features that distinguish variants across approaches (M2). The pipeline introduced by Lopes and Silva~\cite{Lopes2023} was used as the initial conceptual structure for the model, applying top-down modeling to identify differences in the domain (M4). This provided an initial set of high-level stages relevant to microservice identification tooling.

\subsubsection{Bottom-Up Modeling}

In the bottom-up phase (M3), the selected studies were systematically analyzed to extract recurring concepts, dimensions, and decisions reported across monolith-to-micro\-service identification approaches. This analysis was consolidated into a meta-review table, provided in the Appendix~\ref{sec:material}, in which each row corresponds to one of the 11 selected secondary studies and the columns are grouped by the groups of the model, recording the features each study reports. Extracted concepts were iteratively compared and consolidated: when a concept matched an existing feature, it was mapped to that feature; otherwise, a new feature was introduced. Similar or overlapping concepts reported with different terminology across studies were normalized and grouped under a shared feature when they represented the same variability point. The combination of both directions (M5) increased the model's completeness and provided a cross-checking mechanism. To enhance consistency, additional papers were included when they provided more concrete evidence for a specific feature or variation point.

Throughout this process, functional decomposition was used as the criterion for feature-to-subfeature relationships (PP4). Features were designed to represent distinctive functional abstractions (M6), preferring boolean types for comprehension (M10). Feature dependencies were modeled to support end-users in understanding configuration constraints (D2); the expert-only dependency modeling alternative (D1) was therefore not applicable.

The depth of the hierarchy was kept well below the recommended limit of eight levels, with the current model reaching a maximum of five (MO1). Concepts were then organized into mandatory, optional, and alternative features depending on whether they represented common structural elements or distinct variation points, with higher-level groups defined as more abstract AND-groups and lower-level groups using OR-relations to capture specific configuration choices (MO2). To preserve clarity, only simple \textit{requires} and \textit{excludes} cross-tree constraints were permitted (MO4), and high cohesion and low coupling within feature groups were maintained throughout (MO5).

The resulting feature model is presented in Section~\ref{sec:feature-model-structure}.

\subsubsection{Evaluation}
\label{sec:modeling-process-evaluation}

Following the principles of Nešić et al.~\cite{Nesic2019}, the model was then evaluated. The bottom-up phase, where a systematic mapping of representative literature is done, already provides a validation by construction. Additionally, the feature model instantiation was analyzed within the architecture of existing microservice identification tools, as described in Section~\ref{sec:evaluation} (QA2). The tools selected for evaluation were \textit{Mono2Micro}~\cite{Lopes2023}, \textit{Micro2Micro}~\cite{Zhong2025}, and the tools compared by Wang et al.~\cite{Wang2024}. The meta-review table produced in the bottom-up phase served as a reference to verify that model revisions remained consistent with previously established mappings (QA3). Since the modelers are domain experts, formal validation workshops with external experts (QA1) were not required. This paper itself serves as documentation for the features and the obtained feature model (M11).

\subsubsection{Remarks}

Some principles were not applicable given the model's scope and intended use: the introduction of spurious features (M7), default feature values (M8), multiple feature-model views (M9), and splitting the model into sub-models (MO3).

As previously noted, additional papers not captured by the initial Scopus search were identified during the review process. They are detailed in Table~\ref{tab:additional-papers}.

\begin{table*}[htbp]
\caption{Additional papers included outside the formal search}
\label{tab:additional-papers}
\centering
\begin{tabularx}{\textwidth}{@{}l X l l X @{}}
\toprule
\textbf{Reference} & \textbf{Title} & \textbf{Year} & \textbf{Phase(s)} & \textbf{Reason for Inclusion} \\ \midrule
\cite{Lopes2023} & Monolith Microservices Identification: Towards An Extensible Multiple Strategy Tool & 2023 & \makecell{Top-Down \\ Evaluation} & Serves as both the conceptual foundation and primary evaluation target of this work \\
\cite{Nakazawa2018} & Visualization Tool for Designing Microservices with the Monolith-First Approach & 2018 & Bottom-Up & Further supports the modeling of visualization features \\
\cite{Cerny2022} & Microvision: Static analysis-based approach to visualizing microservices in augmented reality & 2022 & Bottom-Up & Further supports the modeling of visualization features \\
\cite{Baresi2017} & Microservices Identification Through Interface Analysis & 2017 & Bottom-Up & Further supports the modeling granularity API endpoint feature \\
\cite{Wang2024} & Microservice Decomposition Techniques: An Independent Tool Comparison & 2024 & Evaluation & Independently compares tools already present in the Scopus results (\cite{Sellami2022}, \cite{Carvalho2020}, \cite{Kalia2021}, \cite{Mazlami2017}) \\
\cite{Zhong2025} & Refactoring Microservices to Microservices in Support of Evolutionary Design & 2025 & Evaluation & Extends the evaluation to microservice-to-microservice refactoring scenarios \\\bottomrule
\end{tabularx}
\end{table*}

\section{Results}
\label{sec:feature-model-structure}

After analyzing the selected studies to answer the research question, we identified several commonalities and variability points across monolith to microservice identification approaches. These were then organized into a feature model that captures the variability of microservice identification and refactoring approaches reported in the literature. It represents the problem space of microservice decomposition by making explicit the design decisions and variation points that characterize existing approaches, enabling systematic comparison and analysis.

\subsection{Feature Model Structure}

To present the proposed feature model clearly, we adopt an incremental approach rather than introducing it all at once. Tracing its construction through the defined modeling process allows us to discuss the rationale behind each design decision that shapes the final model.

\begin{figure}
    \centering
    \includegraphics[width=1\linewidth]{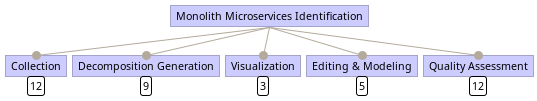}
    \caption{Initial Feature Model - Collapsed}
    \label{fig:initial-fm-collapsed}
\end{figure}

The initial organization of the feature model follows the pipeline (Figure~\ref{fig:pipeline}) proposed by Lopes and Silva~\cite{Lopes2023}, which defines a sequence of stages commonly found in microservice identification approaches. At this level of abstraction, the model can be organized into a set of high-level feature groups corresponding to these stages, namely \textit{Collection}, \textit{Decomposition Generation}, \textit{Visualization}, \textit{Editing \& Modeling}, and \textit{Quality Assessment}. This initial structure provides a simplified and sequential view of the process, serving as a foundation for the top-down phase of the modeling process, as shown in Figure~\ref{fig:initial-fm-collapsed}.

Each of these feature groups serves a distinct goal within the microservice identification process. \textit{Collection} gathers information about the monolithic system under analysis, defining the artifacts and techniques used to produce the input for the subsequent stages. This information feeds \textit{Decomposition Generation}, the core stage, where strategies and algorithms partition the monolith into microservice candidates. The resulting candidates are then made available through \textit{Visualization}, which presents the system and its decompositions to support analysis and interpretation, and \textit{Editing \& Modeling}, which lets users manually adjust a proposed decomposition by splitting and merging microservice candidates or moving entities between them. Finally, \textit{Quality Assessment} evaluates the resulting decomposition, typically through metrics such as cohesion, coupling, and complexity.

Rather than relying on the pipeline alone, we extracted from Lopes and Silva's work~\cite{Lopes2023} the key features that characterize each of these stages, such as the artifacts collected, the decomposition strategies applied, or the metrics computed. Capturing these as features made it possible to populate each group, producing the first extended draft of the feature model, available in full in the Appendix~\ref{sec:material}.

\begin{figure*}
    \centering
    \includegraphics[width=1\linewidth]{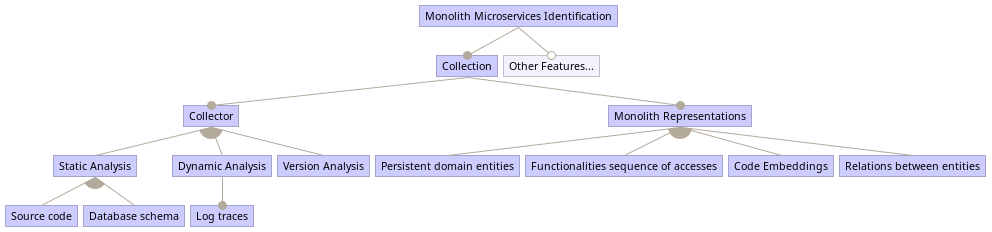}
    \caption{Initial Feature Model - Extended (Collection Feature)}
    \label{fig:initial-fm-collection}
\end{figure*}

Regarding the \textit{Collection} feature, the analysis of Lopes and Silva~\cite{Lopes2023} allows us to distinguish between the techniques used to capture monolith information (\textit{Collector} feature) and their resulting outputs (\textit{Monolith Representation} feature). This distinction is necessary because a single type of output can often be produced by multiple, distinct techniques. For instance, the \textit{Functionalities Sequence of Accesses} can be derived via either \textit{Static Analysis} or \textit{Dynamic Analysis}~\cite{Andrade23}.

Therefore, the \textit{Collector} can use \textit{Static Analysis} techniques to examine the monolithic \textit{Source Code} and \textit{Database Schema}~\cite{Andrade23, Faria23}; \textit{Dynamic Analysis} techniques to gather and evaluate the \textit{Log Traces} of the monolith's execution~\cite{Andrade23}; and \textit{Version History} techniques to track development changes made to the source code~\cite{Lourenco23}.

Applying \textit{Collector} techniques yields various \textit{Monolith Representations}, including: \textit{Persistent Domain Entities}~\cite{Andrade23}, which capture the monolith's persistent data structures; \textit{Functionalities Sequence of Accesses}~\cite{Andrade23}, which map the read and write accesses performed by functionalities on those entities; \textit{Code Embeddings}~\cite{Faria23}, which encode relevant segments of the monolithic source code into vector embeddings; and different types of \textit{Relations between Entities}, such as structural and development relations~\cite{Lourenco23}. Figure~\ref{fig:initial-fm-collection} presents the obtained extended feature model for the \textit{Collection} feature.

\begin{figure}
    \centering
    \includegraphics[width=1\linewidth]{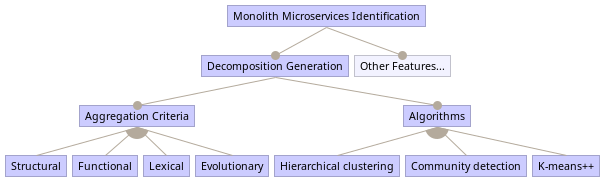}
    \caption{Initial Feature Model - Extended (Decomposition Generation Feature)}
    \label{fig:initial-fm-decomposition}
\end{figure}

These representations are used for \textit{Decomposition Generation}, which Lopes and Silva~\cite{Lopes2023} divide into two parts that we interpret as two different subfeatures: \textit{Aggregation Criteria} and \textit{Algorithms}. Figure~\ref{fig:initial-fm-decomposition} presents its extension. The \textit{Aggregation Criteria} define semantic distances between monolithic elements based on their \textit{Monolith Representations}. These distances are typically measured in one of four ways: structural proximity to enforce modularity; transactional proximity to reduce transactional complexity~\cite{Andrade23}; development proximity to support modularity and team separation~\cite{Lourenco23}; or lexical proximity to enforce business domain aggregation~\cite{Faria23}. All the above works apply \textit{Hierarchical Clustering} for the \textit{Algorithms} feature, but Lopes and Silva~\cite{Lopes2023} mention other algorithms that are also applied.

\begin{figure*}
    \centering
    \includegraphics[width=1\linewidth]{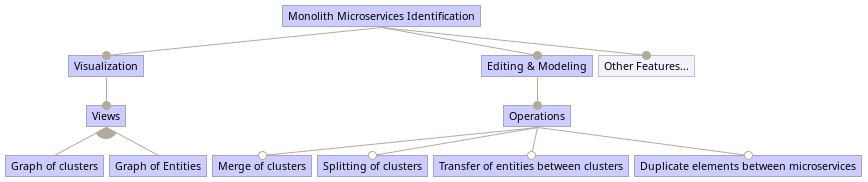}
    \caption{Initial Feature Model - Extended (Visualization and Editing \& Modeling Features)}
    \label{fig:initial-fm-visualization-editing}
\end{figure*}

Figure~\ref{fig:initial-fm-visualization-editing} presents the extension of the initial feature model for the \textit{Visualization} and \textit{Editing \& Modeling} features.

The \textit{Visualization} features utilize a \textit{Graph of clusters} and a \textit{Graph of Entities} to depict the decomposition, leveraging spatial distances and color coding to convey semantic information.
The \textit{Editing \& Modeling} feature enables manual operations on a decomposition, such as merging or splitting clusters.

\begin{figure*}
    \centering
    \includegraphics[width=1\linewidth]{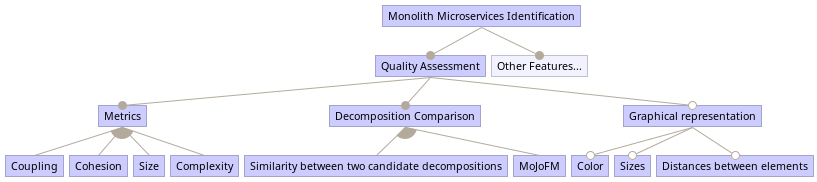}
    \caption{Initial Feature Model - Extended (Quality Assessment Feature)}
    \label{fig:initial-fm-quality}
\end{figure*}

Finally, the \textit{Quality Assessment} feature, whose extension is presented in Figure~\ref{fig:initial-fm-quality}, evaluates the quality of a decomposition using \textit{Metrics} - such as \textit{Coupling}, \textit{Cohesion}, \textit{Size}, and \textit{Complexity}~\cite{Santos20} - and compares different decompositions using MoJoFM~\cite{Zhihua04}. Additionally, it leverages \textit{Colors}, \textit{Sizes}, and \textit{Distances} within a \textit{Graphical Representation} to visually convey these quality indicators.

This extended draft constitutes a first attempt at answering the research question. By treating each stage as an explicit feature group and populating it with concrete features, the model already makes variability explicit: an approach is characterized both by which stages it supports and by which features it selects within them, so that tools relying on different sources, decomposition strategies, or quality metrics, for instance, are distinguished from one another.

To verify whether this initial model captures the diversity of approaches reported in the literature, it was then examined against the surveyed studies and iteratively refined during the bottom-up phase. As new concepts were identified, existing feature groups were reorganized, extended, or redefined, leading to the introduction of additional features and the restructuring of the initial hierarchy. The following sections describe these refinements and motivate the main design decisions that led to the final version of the feature model.

\subsection{Granularity}

In the initial feature model, candidate decompositions consist of groups of domain entities and classes. Even though the information collected from the monolith represents them in different semantic relations, such as structural or dynamic, a candidate microservice remains a group of these types of elements.

However, by applying bottom-up modeling, using selected studies~\cite{Abgaz2023, Abdellatif2021, Mohottige2025, Saucedo2025, Trabelsi2025, Yang2025}, the literature shows that different approaches operate at different levels of abstraction, which made us realize the lack of an explicit representation of the granularity level at which software elements are considered during the decomposition process.

The granularity levels range from fine-grained elements such as methods, e.g.~\cite{Assuncao26}, and classes, e.g.~\cite{desai2021graph}, to coarser-grained elements such as domain entities, e.g.~\cite{Santos22}, functionalities, e.g.~\cite{wei2020feature}, packages, e.g.~\cite{Krause2020}, or API endpoints, e.g.~\cite{ABDULLAH2019}.

Consequently, the initial feature model cannot adequately express this type of feature and its variations. Interestingly, each approach is locked into a specific level of granularity. Because this granularity defines the fundamental building blocks of a microservice, this core design decision significantly shapes the overall nature of each approach.

Therefore, the decision to add a \textit{Granularity} feature group is motivated by the need to explicitly capture this variability, as granularity is a cross-cutting concern that impacts multiple stages, including collection, decomposition, and visualization. This group is mandatory, as every approach must define the level at which decomposition decisions are made.

\begin{figure}
    \centering
    \includegraphics[width=1\linewidth]{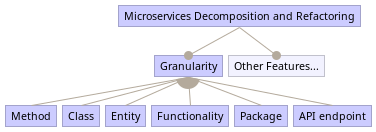}
    \caption{Granularity Feature Group}
    \label{fig:granularity}
\end{figure}

Figure~\ref{fig:granularity} presents the new \textit{Granularity} feature.
The subfeatures of this group are organized as an OR group, allowing multiple granularity levels to be selected simultaneously, as shown in Figure~\ref{fig:granularity}. This design enables approaches to operate at multiple levels at the same time, which allows different strategies that use different granularity levels to coexist within the same tool, which is not the case for existing approaches.

Each of these granularity subfeatures corresponds to different fundamental perspectives on what a microservice is, as shown in Table~\ref{tab:granularity_concepts}.

\begin{table*}[htbp]
\centering
\renewcommand{\arraystretch}{1.4}
\begin{tabular}{|p{0.15\linewidth}|p{0.45\linewidth}|p{0.3\linewidth}|}
\hline
\textbf{Granularity} & \textbf{Goal / Purpose} & \textbf{Microservice Concept} \\ \hline
\textbf{Method} & To extract highly optimized chains of logic by recombining related methods, which can cross traditional class boundaries~\cite{Abgaz2023}. & \textbf{Function-as-a-Service (FaaS)} or highly fine-grained utility functions~\cite{Abgaz2023}. \\ \hline
\textbf{Class} & To partition the system by clustering cohesive and loosely coupled classes together based on structural, semantic, or evolutionary metrics~\cite{Abgaz2023, Abdellatif2021}. & \textbf{Standard Microservice} (a cohesive grouping of related objects)~\cite{Abdellatif2021}. \\ \hline
\textbf{Entity} & To manage and provide access to persistent business data, usually supporting Create, Read, Update, and Delete (CRUD) actions on specific data entities or database tables~\cite{Abdellatif2021}. & \textbf{Entity Service} (Data or Information Service)~\cite{Abdellatif2021}. \\ \hline
\textbf{Functionality} & To group specific business capabilities or use cases, focusing on the business logic rather than structural code boundaries~\cite{Abdellatif2021, Trabelsi2025}. & \textbf{Enterprise-task, Application-task, or Business Service}~\cite{Abdellatif2021}. \\ \hline
\textbf{Package} & To group massive collections of classes and methods without splitting existing source code packages, which lowers the risk of fragmenting the system~\cite{Abgaz2023}. & \textbf{Miniservice or Macroservice}~\cite{Abgaz2023}. \\ \hline
\textbf{API endpoint} & To define and extract service interfaces by analyzing the system's runtime Uniform Resource Identifiers (URIs) or OpenAPI (Swagger) specifications~\cite{Abgaz2023, Trabelsi2025, Baresi2017}. & \textbf{Interface-defined service} (a cohesive group of API operations and associated resources)~\cite{Baresi2017}. \\ \hline
\end{tabular}
\caption{Goals and Associated Microservice Concepts by Granularity Level}
\label{tab:granularity_concepts}
\end{table*}

By explicitly modeling granularity, the feature model enables tools to support multiple decomposition perspectives within the same framework, combine strategies across different levels of granularity, and compare approaches that operate at different conceptual perspectives of what a microservice is, while allowing new approaches to emerge that might propose additional levels of granularity.

\subsection{Representation Collection}

In the initial feature model, the \textit{Collection} feature serves as the interface between the monolith and the decomposition process. It uses data extracted from the monolith to generate the detailed \textit{Monolith Representation} required for decomposition. The extraction process (\textit{Collector}) can vary significantly, depending on the type of data collected and how it is delivered. For instance, it can process source code using a syntactic analysis tool to generate a representation in the form of a call graph, which is then utilized by the decomposition process. Interestingly, the call graph representation can also be derived from the monolith's execution logs. This suggests that the \textit{Collector} feature could likely be split into several finer-grained, cohesive sub-features, separating the source of the collection from the collection technique.

Additionally, the analysis of the literature reveals that many approaches rely on external tools or pre-existing representations rather than implementing their own data collection mechanisms~\cite{Abdellatif2021, Mohottige2025, Yang2025, Trabelsi2025, Saucedo2025}.

\begin{figure*}
    \centering
    \includegraphics[width=1.0\linewidth]{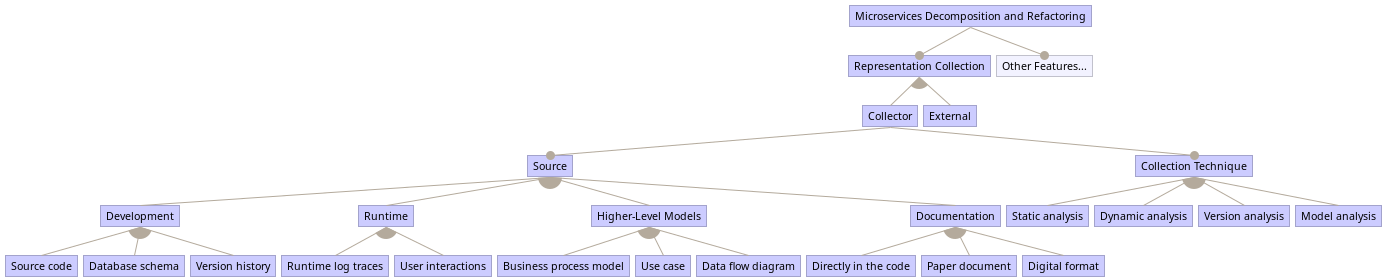}
    \caption{Representation Collection Feature Group}
    \label{fig:representation-collection}
\end{figure*}

Given the breadth of data sources and collection techniques reported across the literature~\cite{Abgaz2023, Abdellatif2021, Mohottige2025, Saucedo2025, Capuano2022, Oumoussa2024, Nassima2025, Oumoussa2025, Trabelsi2025, Yang2025}, the \textit{Collection} group was renamed to \textit{Representation Collection}, and its first sublevel distinguishes between using an inbuilt collector and relying on external sources (see Figure~\ref{fig:representation-collection}). While the main group remains mandatory, the subgroups are within an OR group, allowing for the choice between using a built-in collector, relying on external sources, or combining both options. This design choice accommodates a wide range of scenarios while also allowing for greater flexibility in tool design, as some tools may choose to implement their own collection mechanisms while others may prefer to leverage existing tools and representations.

When a built-in collector is used, the source of the collected data is modeled as an OR group with four categories: \textit{Development}, \textit{Runtime}, \textit{Higher-level Models}, and \textit{Documentation}. The \textit{Development} source focuses on the monolith during its development phase, which includes the version history of the source code. The \textit{Runtime} source captures the monolith during execution, whereas \textit{Higher-level Models} rely on the system's specifications. Additionally, the monolith's \textit{Documentation} can be used to extract the information necessary to generate a representation. Each category includes leaf features representing commonly used artifacts, such as source code, e.g.~\cite{Kamimura2018}, execution traces, e.g.~\cite{jin21}, business models, e.g.~\cite{Daoud2020}, or digital documentation e.g.~\cite{Alahmari2010}.

It is crucial to distinguish between these source types because they yield distinct information, even if they ultimately may generate the same representations~\cite{Abdellatif2021, Saucedo2025, Nassima2025}. So, not all sources offer the same level of detail. For instance, source code and execution logs can both be used to generate the same representation: a call graph. However, this does not imply that these sources contain equivalent information. As shown in~\cite{Andrade23}, the coverage achieved by the two techniques can differ significantly, which reinforces the advantages of this separation. On the other hand, \textit{Documentation} is often criticized for being outdated~\cite{Abdellatif2021}. Additionally, certain source types may be unavailable for a given monolithic system~\cite{Abdellatif2021, Saucedo2025}, forcing the use of less accurate alternatives. Decoupling the source from the representation provides a significant advantage here: it offers the flexibility to handle missing sources. If a specific source is unavailable, the same representation can still be generated using an alternative, albeit with varying degrees of accuracy.

Additionally, the model separates data sources from collection techniques, as the same technique may be applied to different sources. These collection techniques are responsible for extracting information from the selected sources and producing representations that can be used in subsequent stages of the microservice identification process. The techniques are modeled as an OR group with four options: \textit{Static Analysis}, e.g.,~\cite{Baresi2017}, which extracts information from source code, database schemas, or documentation; \textit{Dynamic Analysis}, e.g., ~\cite{jin21}, which collects information from runtime artifacts; \textit{Version Analysis}, e.g.~\cite{Mazlami2017}, which mines software configuration management repositories to capture evolutionary aspects of the system; and \textit{Model Analysis}, e.g.,~\cite{gysel2016service}, which operates on higher-level representations.

The overall structure of the \textit{Representation Collection} feature group is illustrated in Figure~\ref{fig:representation-collection}. The leaf features under both data sources and collection techniques represent commonly used artifacts and methods reported in the literature. They are designed to be extensible, allowing new sources and techniques to be added as the state of the art evolves.

There are dependencies between these features, in the sense that some \textit{Collection Techniques} require some particular \textit{Sources}. Table~\ref{tab:collection-technique-dependencies} summarizes the dependencies between collection techniques and the data sources they require.

\begin{table*}[htbp]
\caption{Constraints between collection techniques and data sources}
\label{tab:collection-technique-dependencies}
\centering
\begin{tabularx}{\textwidth}{@{}l X @{}}
\toprule
\textbf{Feature} & \textbf{Constraint} \\ \midrule
CollectionTechnique.StaticAnalysis & \textbf{Requires} Source.Development.(SourceCode \textbar{} DatabaseSchema) \\
CollectionTechnique.DynamicAnalysis & \textbf{Requires} Source.Runtime \\
CollectionTechnique.VersionAnalysis & \textbf{Requires} Source.Development.VersionHistory \\
CollectionTechnique.ModelAnalysis & \textbf{Requires} Source.Higher-LevelModels \textbar{} Source.Documentation\\\bottomrule
\end{tabularx}
\end{table*}

\textit{Static Analysis} techniques require structured source code and depend heavily on the specific programming language and technology stack. In contrast, \textit{Dynamic Analysis} is language- and framework-agnostic, relying instead on traces generated during the monolith's runtime execution (such as logs) to infer its representation. \textit{Version Analysis} deduces this architecture from the source code's development history, while \textit{Model Analysis} leverages the monolith's high-level models and documentation and is frequently based on natural language processing techniques.

\subsection{Representation}
\label{sec:representation}

Following the separation between representation collection and monolith representation, the initial feature model presents some types of representation, such as the sequences of accesses done by the monolith functionalities. It is now necessary to enhance the feature model by a bottom-up analysis of the variety of representation strategies reported in the literature~\cite{Abgaz2023, Mohottige2025, Trabelsi2025, Yang2025, Zhong2025}. The new \textit{Representation} feature group captures the different types of semantically rich artifacts, distilled from the monolithic source by the collection techniques, used to feed the decomposition decisions.

\begin{figure*}
    \centering
    \includegraphics[width=1.0\linewidth]{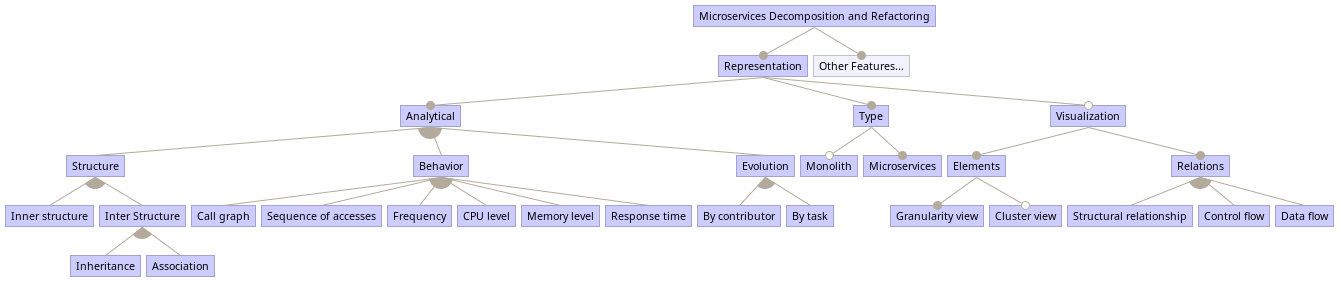}
    \caption{Representation Feature Group}
    \label{fig:representation}
\end{figure*}

The \textit{Representation} feature group, shown in Figure~\ref{fig:representation}, defines how system information is structured and made accessible for analysis and visualization.

This group is mandatory, as every approach must define some form of representation to support decomposition decisions, and it is composed of three main feature subgroups: \textit{Analytical}, \textit{Type}, and \textit{Visualization}.

The \textit{Analytical} feature is the natural evolution of the \textit{Monolith Representations} feature of the initial feature model. It captures the relevant elements and relations among them using a syntactic representation that can be used to model the system. These representations, like the different perspectives of a software architecture, are divided into subfeatures that view the monolith as a code artifact, \textit{Structure} subfeature, a running artifact, \textit{Behavior} subfeature, or a development artifact, \textit{Evolution} subfeature.

\textit{Structure} representations describe how monolith code elements are organized. They can capture inner structure, such as the internal composition of elements, as well as inter-structure relationships, including inheritance and association dependencies~\cite{Levezinho25}. These representations can be obtained using static analysis techniques or model analysis techniques.

\textit{Behavior} representations characterize runtime interactions between system elements. Call graphs, e.g.~\cite{Matias2020}, sequences of accesses, e.g.~\cite{Nunes19}, and resource-related information, e.g.~\cite{Zhang2020}, such as CPU usage or response time. Each of those behavior representations can be obtained using static, dynamic, or model analysis.

Finally, \textit{Evolution} representations capture how the system changes over time and are obtained through version analysis techniques~\cite{Mazlami2017}. They include information such as co-changing contributors (\textit{By Contributor}), co-changing files (\textit{By Task}), or other evolution metrics that might be useful to guide decomposition decisions. These representations can be obtained through \textit{Version Analysis}.

The \textit{Analytical} subgroup is modeled as an OR-group, allowing multiple types of representations to be combined within the same approach, thus supporting richer and more comprehensive system characterizations.

In Table~\ref{tab:analytical-dependencies} are presented the dependencies between the \textit{Analytical} representation features and the collection techniques that can produce them when using an inbuilt collector.

\begin{table*}[htbp]
\caption{Constraints between analytical representation features and collection techniques, when their collection is not \textit{External}}
\label{tab:analytical-dependencies}
\centering
\begin{tabularx}{\textwidth}{@{}l X @{}}
\toprule
\textbf{Feature} & \textbf{Constraint} \\ \midrule
Analytical.Structure & \textbf{Requires} CollectionTechnique.(StaticAnalysis \textbar{} ModelAnalysis) \\ \midrule
Analytical.Behavior & \\
\hspace{3mm} .CallGraph & \textbf{Requires} CollectionTechnique.(StaticAnalysis \textbar{} DynamicAnalysis \textbar{} ModelAnalysis) \\
\hspace{3mm} .SequenceOfAccesses & \textbf{Requires} CollectionTechnique.(StaticAnalysis \textbar{} DynamicAnalysis \textbar{} ModelAnalysis) \\
\hspace{3mm} .Frequency & \textbf{Requires} CollectionTechnique.DynamicAnalysis \\
\hspace{3mm} .CPULevel & \textbf{Requires} CollectionTechnique.DynamicAnalysis \\
\hspace{3mm} .MemoryLevel & \textbf{Requires} CollectionTechnique.DynamicAnalysis \\
\hspace{3mm} .ResponseTime & \textbf{Requires} CollectionTechnique.DynamicAnalysis \\ \midrule
Analytical.Evolution & \textbf{Requires} CollectionTechnique.VersionAnalysis \\ \bottomrule
\end{tabularx}
\end{table*}

Recent approaches have shifted focus from extracting microservices from monolithic architectures to the broader objective of microservice refactoring, where existing services are optimized, split, and merged~\cite{Zhong2025, Yang2025}.

Since the core goals and techniques are the same, the initial feature model should be extended to support both scenarios. Therefore, it explicitly distinguishes between monolithic and microservice systems by adding the \textit{Type} feature group under the \textit{Representation} feature group, accommodating decomposition, refactoring, and optimization strategies to both monolith and microservice systems.

Within this group, the \textit{Microservices} representation is mandatory, as every approach yields it as the outcome of decomposition or refactoring. Conversely, the \textit{Monolith} representation is optional, since microservice-to-microservice scenarios do not inherently require a monolithic representation.

Both the \textit{Monolith} and the \textit{Microservices} representations can be obtained from internal or external collectors, but the \textit{Microservices} representations additionally participate in the refactoring loop since they both feed into and result from refactoring features. \textit{Analytical} representations can be generated using specific microservice system collectors. However, candidate microservice representations must also contain the additional information that groups these elements. For instance, a \textit{Call Graph} should clearly distinguish between intra-candidate and inter-candidate microservice calls.

In the initial feature model, visualization was considered a top-level feature because its purpose was to display the microservice system resulting from decomposition. Since we are now integrating representations for both the monolith and microservices, it makes sense to classify \textit{Visualization} as a subfeature of \textit{Representation}, making it applicable to both \textit{Monolith} and \textit{Microservices} types. However, as most approaches do not provide any visualization, the \textit{Visualization} feature is optional, but its inclusion allows for a better interpretation of the decomposition and refactoring results.

The extended feature model distinguishes between two complementary aspects of visualization: the elements to be visualized and the types of relationships between them. 

Visualization elements define how this information is presented and are independent of the underlying representation, as long as the required data is available. This subgroup is divided into granularity view and cluster view: the former ensures consistency with the \textit{Granularity} feature selected, presenting elements at the same level of abstraction, while the latter enables the representation of groups of elements, which is particularly relevant when analyzing or refining decompositions. The latter is optional, as not all approaches explicitly represent clusters, in which case the microservices are implicit by the use of colors or proximity between elements. The relations permit highlighting semantic relationships between the elements, such that two classes were changed by the same developer.

The \textit{Visualization} subgroup is modeled as an AND-group because it is necessary to define the elements to be visualized and the relationships to be visualized. It is also designed to be extensible, with leaf features being in an OR-group, allowing multiple visualization mechanisms to coexist within the same tool.

Visualization is inherently dependent on the underlying source, as different sources, such as profiled data from a monolithic prototype or static analysis of microservice codebases, lead to different visualizations~\cite{Nakazawa2018, Cerny2022}.
Relationship types represent different perspectives on the system and require specific representations.

\begin{table*}[htbp]
\caption{Constraints between visualization relation types and analytical features}
\label{tab:visualization-deps}
\centering
\begin{tabularx}{\textwidth}{@{}l X @{}}
\toprule
\textbf{Feature} & \textbf{Constraint} \\ \midrule
Visualization.Relations.StructuralRelationship & \textbf{Requires} Analytical.(Structure \textbar{} Evolution) \\
Visualization.Relations.ControlFlow & \textbf{Requires} Analytical.(Structure \textbar{} Behavior) \\
Visualization.Relations.DataFlow & \textbf{Requires} Analytical.(Structure \textbar{} Behavior) \\\bottomrule
\end{tabularx}
\end{table*}

Table~\ref{tab:visualization-deps} presents the dependencies between \textit{Visualization} relation types and the analytical representations they require.

Although the research question that guided this work targets the identification of microservices in monolithic systems, the introduction of the \textit{Microservices} representation extends the model also to cover the refactoring of existing microservice systems. The research question is therefore updated to "How can we identify and represent the common parts and variability of the approaches for the identification and refactoring of microservices in monolithic and microservice systems?".

\subsection{Refactoring}

In the top-down phase of the feature modeling process, we identified two key features: \textit{Decomposition Generation}, which generates microservice candidates, and \textit{Editing \& Modeling}, which lets users manually adjust them. The latter ultimately serve to optimize a specific decomposition, a goal that, as mentioned in the previous section, recent research also addresses via automated refactoring approaches~\cite{Zhong2025, Yang2025}.

Therefore, both refactoring and edit-and-modeling operations share the same purpose, though the former relies on automated techniques while the latter leverages human expertise. Furthermore, the data representations used by these decomposition and refactoring techniques are highly similar, as detailed in the \textit{Representation} feature, defined in the previous section.

To address this, and also driven by the analysis of approaches reported across the literature~\cite{Abgaz2023, Abdellatif2021, Mohottige2025, Saucedo2025, Oumoussa2024, Nassima2025, Oumoussa2025, Trabelsi2025, Yang2025, Lopes2023, Zhong2025}, these two groups are then combined into a single \textit{Refactoring} group that captures the strategies and techniques used to decompose monolithic systems into microservices and to refactor microservice systems. The manual editing can then be seen as the counterpart of automatic algorithms. 

\begin{figure*}
    \centering
    \includegraphics[width=1\linewidth]{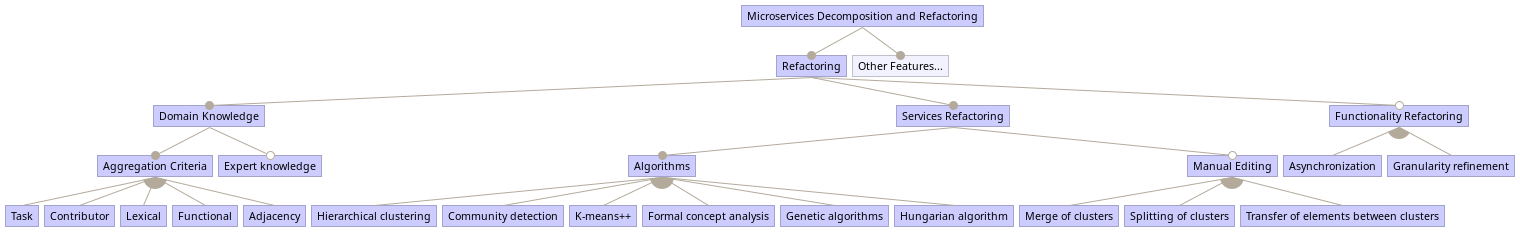}
    \caption{Refactoring Feature Group}
    \label{fig:refactoring}
\end{figure*}

This \textit{Refactoring} group, which can be seen in Figure~\ref{fig:refactoring}, is mandatory, as every approach must define a refactoring strategy. Still, its subgroups are designed to be extensible, allowing new strategies and techniques to be added as the state of the art evolves.

All algorithms require some form of aggregation criteria to group system elements during decomposition or optimization. This requirement is captured by the mandatory \textit{Aggregation Criteria} subgroup within \textit{Domain Knowledge}. The \textit{Aggregation Criteria} feature was already identified in the initial feature model, but it is now a subfeature of \textit{Domain Knowledge} such that the model also includes \textit{Expert knowledge} as input for the refactoring activities. The aggregation criteria leaf features represented are \textit{Task}, \textit{Contributor}, \textit{Lexical}, \textit{Functional} and \textit{Adjacency}.

\textit{Task} and \textit{Contributor} criteria, e.g.~\cite{Mazlami2017}, rely on evolution by task and evolution by contributor information, respectively, and aggregate elements when they were changed in the same tasks or by the same developer. They result from a refinement of the \textit{Evolutionary} feature in the initial feature model. \textit{Lexical}, e.g.~\cite{Al-Debagy2021}, is derived from representations that can be tokenized, such as structural representations, call graphs, or sequences of accesses, which aggregate elements when they exhibit lexical similarities, such as classes and methods. \textit{Functional}, e.g.~\cite{jin21}, exploits behavior information, such as call graphs or sequences of accesses, to aggregate elements that are shared by the same functionalities of the systems. \textit{Adjacency}, e.g.~\cite{adjoyan2014service}, is based on structural or behavioral representations and aggregates elements that are tightly coupled through frequent direct interactions, such as method calls or shared data accesses. The \textit{Adjacency} feature is an extension of the \textit{Structural} feature in the initial feature model, where non-structural relations between elements, such as invocations, are also considered.

As multiple algorithms can be applied to the same system, this subgroup is also modeled as an OR group, allowing multiple criteria to be combined within the same refactoring process, and even extending the model with new aggregation criteria as they are proposed in the literature.

\begin{table*}[htbp]
\caption{Constraints between aggregation criteria and representation features}
\label{tab:aggregation-dependencies}
\centering
\begin{tabularx}{\textwidth}{@{}l X @{}}
\toprule
\textbf{Feature} & \textbf{Constraint} \\ \midrule
AggregationCriteria.Task & \textbf{Requires} Analytical.Evolution.ByTask \\
AggregationCriteria.Contributor & \textbf{Requires} Analytical.Evolution.ByContributor \\
AggregationCriteria.Lexical & \textbf{Requires} Analytical.Structure \textbar{} Analytical.Behavior.(CallGraph \textbar{} SequenceOfAccesses) \\
AggregationCriteria.Functional & \textbf{Requires} Analytical.Behavior.(CallGraph \textbar{} SequenceOfAccesses) \\
AggregationCriteria.Adjacency & \textbf{Requires} Analytical.(Structure \textbar{} Behavior) \\\bottomrule
\end{tabularx}
\end{table*}

Table~\ref{tab:aggregation-dependencies} summarizes the dependencies between \textit{Aggregation Criteria} and the \textit{Representation} features they require. The \textit{Aggregation Criteria} extract semantic information from the \textit{Representation} that is used by the refactoring algorithms.

As already mentioned, the \textit{Domain Knowledge} subgroup also includes \textit{Expert knowledge}, which allows architects to introduce constraints or preferences into the refactoring process, such as limiting the maximum number of microservices or defining that two elements have to belong to the same microservice, e.g.,~\cite{Zhong2025}. This feature is optional, as not all algorithms support the inclusion of external constraints, and tools may still achieve their main refactoring objectives without implementing this capability.

By combining the goals of monolith decomposition and microservice refactoring, the initial model's \textit{Algorithms} and \textit{Editing \& Modeling} features are now grouped under \textit{Services Refactoring}.

Therefore, the \textit{Services Refactoring} subgroup captures the strategies that can be applied to perform the refactoring of a system. It includes a mandatory \textit{Algorithms} subgroup, since the goal of the proposed model is to support the automation of the refactoring process, and so at least one algorithm must be implemented.

The model currently includes \textit{Hierarchical Clustering}, e.g.~\cite{Al-Debagy2021}, \textit{Community Detection}, e.g.~\cite{Matias2020}, \textit{K-means++}, e.g.~\cite{ABDULLAH2019}, \textit{Formal Concept Analysis}, e.g.~\cite{mostefai2012locating}, \textit{Genetic Algorithms}, e.g.~\cite{Zhang2020}, and the \textit{Hungarian Algorithm}, e.g.~\cite{Chen2017}, reflecting commonly used strategies in the literature. Each of these algorithms may be applied to monolithic systems, to microservice architectures that require modernization, or to systems that have already been decomposed into microservice architectures.

The \textit{Algorithms} leaf features are organized as an OR group, allowing multiple algorithms to be applied within the same refactoring process, either in parallel, to compare their results, or sequentially, to optimize the results of one algorithm with another. For instance, a tool may apply a hierarchical clustering algorithm to generate an initial partitioning of microservice candidates, and then a genetic algorithm to optimize this partitioning by merging or splitting candidates based on specific quality metrics.

The model is designed to be extensible, explicitly encouraging the evaluation and integration of new techniques as they are proposed in the literature. None of the currently represented algorithms imposes additional constraints, but new techniques may have different characteristics and requirements, which can lead to the definition of new constraints. For instance, an algorithm may depend on a specific aggregation criterion or might not allow reconciliation with expert knowledge, which would require the definition of an \textit{exclude} constraint.

The optional \textit{Manual Editing} subgroup within the \textit{Services Refactoring} group, previously \textit{Editing \& Modeling} feature in the initial feature model, allows users to manually adjust the proposed decomposition by merging or splitting microservice candidates or transferring elements between clusters, e.g.~\cite{Nakazawa2018}. This subgroup is optional, as not all tools provide manual editing capabilities, but its inclusion can significantly enhance usability and allow architects to incorporate their domain knowledge and preferences into the final decomposition. If implemented together with \textit{Expert knowledge} from the \textit{Domain Knowledge} subgroup, manual editing can also provide a powerful mechanism for architects to iteratively refine the decomposition by applying constraints and adjustments based on their expertise.

\begin{table*}[htbp]
\caption{Constraints between manual editing actions and visualization features}
\label{tab:manual-editing-deps}
\centering
\begin{tabularx}{\textwidth}{@{}l X @{}}
\toprule
\textbf{Feature} & \textbf{Constraint} \\ \midrule
ManualEditing.MergeOfClusters & \textbf{Requires} Visualization.VisualizationElements.ClusterView \\
ManualEditing.SplittingOfClusters & \textbf{Requires} Visualization.VisualizationElements.ClusterView \\
ManualEditing.TransferOfElementsBetweenClusters & \textbf{Requires} Visualization.VisualizationElements.ClusterView \\\bottomrule
\end{tabularx}
\end{table*}

Table~\ref{tab:manual-editing-deps} presents the dependencies between \textit{Manual Editing} subfeatures and \textit{Visualization} subfeatures. The \textit{Cluster View} is required for manipulations that change the candidate microservices, which are visually represented as clusters. Depending on the type of visualization, other operations and their corresponding dependencies may be defined.

Finally, the \textit{Refactoring} group also includes the optional \textit{Functionality Refactoring} subgroup. This subgroup captures functionality refactorings that go beyond structural reorganization, such as splitting functionalities into smaller units or introducing asynchronous communication between components that become separated into different microservices~\cite{Lopes2023, Zhong2025}. Although functionality refactoring is less frequently addressed in the literature, it naturally arises during microservice extraction and evolution and is essential for modeling realistic refactoring scenarios. It is an optional feature, as not all tools support this kind of refactoring, nor do they need it to achieve their goals. However, its inclusion highlights the extensibility of the model and enables future tools to support additional functionality refactoring strategies. Almeida and Silva~\cite{Almeida20} and Correia and Silva~\cite{Correia2022} propose manual and automatic approaches, respectively, to reduce system complexity within a candidate decomposition. Both methods achieve this by refactoring fine-grained inter-microservice invocations into coarse-grained ones.

\subsection{Quality Assessment}

The \textit{Quality Assessment} feature group, identified during the top-down phase, remained largely stable in contrast to other parts of the initial feature model, and its purpose can easily be extended to include the evaluation of the quality of refactorings. Moreover, the surveyed literature consistently emphasizes the importance of evaluating decomposition quality using metrics such as cohesion, coupling, and complexity~\cite{Abgaz2023,Abdellatif2021, Mohottige2025, Saucedo2025, Capuano2022, Oumoussa2024, Auer2021, Oumoussa2025, Trabelsi2025}.

This group is mandatory, as well as its \textit{Metrics} subgroup, as assessing the quality of candidate microservice decompositions is essential for both researchers and practitioners.

The \textit{Metrics} subgroup aggregates quantitative measures used to evaluate decomposition quality. For the evaluation of a single decomposition, the only mandatory metric captures the size of the elements composing the decomposition and can always be computed independently of the data sources, representations, or decomposition strategies used, e.g.~\cite{Taibi2020}. Other metrics, such as \textit{Cohesion}, e.g.~\cite{jin21}, \textit{Coupling}, e.g.~\cite{jin21}, \textit{Complexity}, e.g.~\cite{Santos20}, and \textit{Team Size Reduction}, e.g.~\cite{Mazlami2017}, are optional, as their applicability depends on the available source information and the specific decomposition approach adopted. This subgroup is modeled as an OR group, allowing multiple metrics to be combined to provide a more comprehensive assessment of decomposition quality.

To support the comparison of different candidate microservices decompositions, the model requires the implementation of metrics like MoJoFM~\cite{Zhihua04} whenever comparative evaluation is supported, as it is one of the most widely adopted metrics in the literature for assessing similarity between decompositions. Additional comparison metrics may also be supported, as well as comparisons based on single-decomposition metric values, such as the level of cohesion of a candidate decomposition.

Finally, the \textit{Quality Assessment group} also includes the optional \textit{Graphical Comparison} subgroup, which leverages visualization techniques to compare decompositions. Visual cues such as color, size, spatial distance, and other graphical elements can help users identify similarities and differences between candidate decompositions, assisting analysis and decision-making, e.g.~\cite{Nakazawa2018}. Although optional, graphical comparison capabilities can significantly enhance usability and interpretability when available.

\begin{table*}[htbp]
\caption{Constraints between quality assessment metrics and analytical representation features}
\label{tab:metrics-dependencies}
\centering
\begin{tabularx}{\textwidth}{@{}l X @{}}
\toprule
\textbf{Feature} & \textbf{Constraint} \\ \midrule
Metrics.SingleDecomposition.Coupling & \textbf{Requires} Analytical.(Structure \textbar{} Behavior) \\
Metrics.SingleDecomposition.Cohesion & \textbf{Requires} Analytical.(Structure \textbar{} Behavior) \\
Metrics.SingleDecomposition.Complexity & \textbf{Requires} Analytical.(Structure \textbar{} Behavior) \\
Metrics.SingleDecomposition.TeamSize & \textbf{Requires} Analytical.Evolution.ByContributor \\\bottomrule
\end{tabularx}
\end{table*}

Table~\ref{tab:metrics-dependencies} presents the dependencies between quality assessment metrics and the analytical representation features. Note that, since the same metric can be calculated using different types of information, it is not clear that approaches that use the same metric name are actually measuring the same thing. Therefore, the feature model can also help to highlight these possible inconsistencies between the approaches' evaluations.

\begin{figure*}
    \centering
    \includegraphics[width=0.6\linewidth]{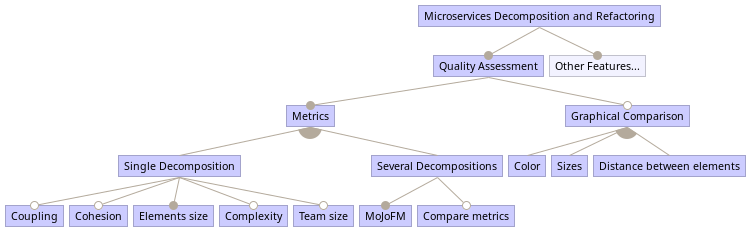}
    \caption{Quality Assessment Feature Group}
    \label{fig:qualityAssessment}
\end{figure*}

Figure~\ref{fig:qualityAssessment} presents the \textit{Quality Assessment} feature group.

Overall, \textit{Quality Assessment} is mandatory, as evaluating the quality of decompositions is essential for both practitioners and researchers. Its subgroups are designed to be extensible, allowing new metrics and comparison techniques to be incorporated as the state of the art evolves.

\subsection{Final Feature Model Overview}

The final feature model integrates all the refinements discussed above into a unified structure. It consists of five main feature groups: \textit{Granularity}, \textit{Representation Collection}, \textit{Representation}, \textit{Refactoring}, and \textit{Quality Assessment}.

Each group captures a specific aspect of the microservice identification and refactoring process, while their combination enables the representation of a wide range of approaches reported in the literature.

\subsection{Cross-tree Constraints}

To accurately capture the variability of microservice identification and refactoring approaches, the proposed feature model includes several cross-tree constraints that express dependencies between features that cannot be fully represented through the hierarchical structure alone. Following Nešić et al.~\cite{Nesic2019}, and to preserve clarity, only simple cross-tree constraints are used, which in the current model are all \textit{requires} dependencies.

These constraints were already introduced in the previous sections as part of the model refinement process and are summarized in the tables presented throughout the text (Tables~\ref{tab:collection-technique-dependencies}, \ref{tab:analytical-dependencies}, \ref{tab:visualization-deps}, \ref{tab:aggregation-dependencies}, \ref{tab:manual-editing-deps}, and \ref{tab:metrics-dependencies}). However, as new approaches are incorporated and additional features are introduced, new cross-tree constraints may be required to accurately capture their dependencies and exclusions.

\begin{figure}
\centering
\includegraphics[width=1\linewidth]{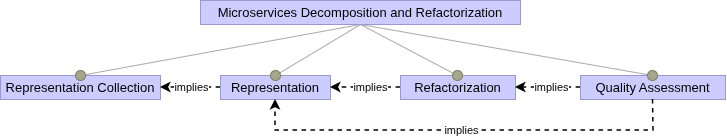}
\caption{Cross-tree Constraints - Abstraction}
\label{fig:feature-model-constraints_abstraction}
\end{figure}

An abstracted view of the constraints is presented in Figure~\ref{fig:feature-model-constraints_abstraction}, where only the main feature groups are represented and dashed arrows indicate a \textit{requires} dependency between the connected groups. This abstraction closely resembles an inverse view of the microservice identification pipeline proposed by Lopes and Silva~\cite{Lopes2023} (Figure~\ref{fig:pipeline}).

The \textit{requires} relationships indicate that a feature can only be selected if the features it depends on are also selected. Consequently, the direction of the arrows reflects the temporal sequence of stages in the microservice identification pipeline, highlighting how the proposed feature model provides a formal, variability-aware counterpart to the pipeline implemented by tools such as Mono2Micro~\cite{Lopes2023}.

\section{Evaluation}
\label{sec:evaluation}

\begin{table*}[htbp]
\caption{Sources and Tools Selected for evaluation}
\label{tab:evaluation-tools}
\centering
\begin{tabularx}{\textwidth}{@{}l l X @{}}
\toprule
\textbf{Tool} & \textbf{Source} & \textbf{Included/Excluded reason} \\ \midrule
Mono2Micro & \cite{Lopes2023} & Included using \texttt{codebase-map} \\
Mono2Micro (IBM) \cite{Mono2Micro_IBM_2020} & \cite{Wang2024} & Included using \texttt{docs-map} \\
HyDec \cite{Sellami2022} & \cite{Wang2024} & Included using \texttt{codebase-map} \\
Data-Centric \cite{Romani22} & \cite{Wang2024} & Excluded, no online codebase or documentation \\
Log2MS \cite{Log2MS} & \cite{Wang2024} & Excluded, no online codebase or documentation \\
MEM \cite{Mazlami2017} & \cite{Wang2024} & Included using \texttt{codebase-map} \\
CARGO \cite{CARGO} & \cite{Wang2024} & Included using \texttt{codebase-map} \\
MOSAIC \cite{Filippone2023} & \cite{Wang2024} & Included using \texttt{codebase-map} \\
Micro2Micro & \cite{Zhong2025} & Included using \texttt{codebase-map} \\\bottomrule
\end{tabularx}
\end{table*}

As described in Section~\ref{sec:modeling-process-evaluation}, the feature model is evaluated by analyzing its instantiation within the architecture of existing microservice identification tools. The selected tools were \textit{Mono2Micro}~\cite{Lopes2023}, \textit{Micro2Micro}~\cite{Zhong2025}, and the tools compared by Wang et al.~\cite{Wang2024}. However, not all of them could be analyzed: tools without a publicly available codebase or documentation were excluded. Table~\ref{tab:evaluation-tools} lists the tools considered, indicating which were included and the reason for excluding the remaining ones.

To ensure that the analysis of these tools was systematic and that their results were reproducible, including when mapping other tools or when the feature model was later modified, extended, or improved, the mapping was supported by a set of Claude\footnote{\href{https://claude.ai}{https://claude.ai}} skills that carry out the mapping automatically. The design, refinement, and use of these skills are described in the next section.

\subsection{Evaluation Methodology}
\label{sec:evaluation-methodology}

\begin{figure}
    \centering
    \includegraphics[width=0.5\linewidth]{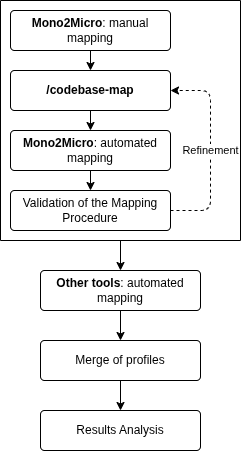}
    \caption{Evaluation Methodology}
    \label{fig:evaluation-methodology}
\end{figure}

Figure~\ref{fig:evaluation-methodology} illustrates the evaluation methodology, which started with a manual mapping of the features of \textit{Mono2Micro}~\cite{Lopes2023}. This tool was chosen as the first evaluation target because the top-down phase of the feature modeling process was based on its analysis, and a more straightforward mapping was therefore expected. This strategy also allowed us to calibrate the Claude skills.

\begin{table*}[htbp]
\caption{Meaning of colors in the mapping}
\label{tab:color-meaning}
\centering
\begin{tabularx}{\textwidth}{@{}l X X @{}}
\toprule
\textbf{Color} & \textbf{Meaning} & \textbf{Notes} \\ \midrule
Dark Green & Every feature in the group was mapped to the tool & Only used for feature groups, never for leaf features \\
Light Green & The feature was mapped to the tool &  \\
Yellow & The feature was not mapped to the tool, but it is optional &  \\
Red & The feature was not mapped to the tool, although it was mandatory &  \\
Orange & The feature was not mapped to the tool and this violates a constraint &  \\
Cyan & Feature present in the tool but absent from the model & Only used in manual mapping \\
Magenta & The mapping of this feature is inconclusive & Only used in automated mapping \\\bottomrule
\end{tabularx}
\end{table*}

The mapping was performed by coloring the features of the feature model according to Table~\ref{tab:color-meaning}. Because the model is intended to be extensible and most of its leaf features are examples of possible implementations, individual leaf features are never colored dark green. This color is therefore reserved for feature groups in which every proposed leaf feature was mapped.

Features present in the tool codebase that could not be directly mapped to the model were also identified and, where possible, added to the model, marked with a cyan color. For \textit{Mono2Micro}, no such feature was found that could not be incorporated, indicating that the model captures the necessary feature groups and remains extensible.

The results of this mapping are available in the Appendix~\ref{sec:material} and are discussed further in Section~\ref{sec:evaluation-results}.

\subsubsection{Codebase-Map}
\label{sec:skill-codebase-map}

To automate the mapping of tools onto the feature model, the \texttt{codebase-map} skill was developed. Given the codebase of a tool, the skill produces three artifacts: a human-readable report, an analysis document, and a color profile.

The report's goal is to support a straightforward manual verification of the mapping. It collects pointers to all the material relevant to the tool, namely the other artifacts produced by the skill, the tool's codebase, and, when available, the paper that introduces the tool. It also documents how to set up and run the tool and records any gotchas encountered during the analysis.

The analysis document is the core artifact and serves both the skill itself and the user, who can use it to validate the results and trace their supporting evidence. Its structure mirrors the feature model: each section corresponds to a feature group and is populated with one bullet per leaf feature, stating whether that feature could be mapped to the tool. To reduce errors and allow the user to independently verify the mapping, every finding must cite at least one \texttt{file:line} reference from the actual codebase. This detailed analysis is then condensed into a summary table that reports, for each feature, whether it is supported by the tool and the evidence for that decision.

The analysis document closes with an observations section that captures insights that fall outside the per-feature mapping. Mirroring how the manual mapping surfaced additional leaves that could be added to the feature model, this section records \textbf{model gap candidates}: features that are absent from the model but could be easily included. Features that are present in the tool and yet too specific to fit as leaves under an existing group are instead listed under \textbf{other features outside the model}. Dimensions where the model's vocabulary does not quite match the tool's design, such as a concept mismatch or a partial overlap, are recorded in a \textbf{model vocabulary fit} section, whereas \textbf{cross-tree constraint violations} are reported separately. Finally, any \textbf{noteworthy design decisions} observed in the tool are also registered.

The color profile is the machine-readable counterpart of the manual colored mapping. It assigns every feature of the model a color according to the meaning defined in Table~\ref{tab:color-meaning}, so that the outcome of the mapping can be rendered directly over the feature model and inspected visually.

This skill was refined iteratively until it produced stable results across runs that were consistent with the manual mapping. This process led to the introduction of rules such as the requirement to justify every finding with evidence and the preference for inspecting the code directly rather than relying solely on the documentation shipped with the codebase.

\paragraph{Mapping Validation}
\label{sec:mapping-validation}

To assess the skill's reliability, the mapping it produced for \textit{Mono2Micro} was compared against the manual mapping described earlier in this section. The two mappings agreed on the vast majority of features, including all of the model gap candidates, given that the skill independently surfaced the same four leaves that had been added to the model during the manual mapping. They diverged on only three features, discussed below.

The first divergence concerned the \textit{Inner structure}, marked \textit{Magenta} by the skill but \textit{Light Green} in the manual mapping. As defined in Table~\ref{tab:color-meaning}, \textit{Magenta} signals that the mapping is inconclusive and leaves the feature for the user to decide. Therefore, this divergence reflects the skill deferring judgment rather than making a claim. Its reasoning was that, although the tool reads the inner structure, it uses it only to compute association and inheritance weights between entities and never to generate the microservices themselves, leaving the feature's status genuinely ambiguous. On review, the manual mapping proved wrong: \textit{Inner structure} is not, in fact, a feature of the tool.

This case illustrates the skill's ability to defer when evidence is thin, which is a desirable property since it avoids false positives. At the same time, it pointed out a human error in the manual mapping, which is a common risk when the analyst is unfamiliar with the tool's codebase, design, or terminology. The skill's ability to cite evidence for every finding also makes it easier for the user to verify the mapping and correct mistakes.

The \textit{Distance between elements} feature was marked \textit{Magenta} by the tool and \textit{Light Green} by the manual mapping, but this one was resolved differently. The skill analysis \textit{Distance between elements} argued that the feature was present only as the default behavior of the algorithm used to visualize the decomposition, rather than being explicitly implemented by the tool. That observation was accurate but immaterial: how a feature comes to be supported does not change the fact that the tool supports it, so the manual \textit{Light Green} mapping was correct.

The remaining case, \textit{Granularity Refinement} feature, was mapped \textit{Light Green} manually but \textit{Yellow} by the skill. The skill's reasoning again turned on explicitness, since the feature is implemented by the same script that handles \textit{Asynchronization} rather than in dedicated modular code. As in the previous case, the manual mapping was correct, and the skill should have marked the feature at least \textit{Magenta} rather than dismissing it as an unmapped option.

This example highlights a deliberate design decision. We could have added a rule instructing the skill to mark any feature present in the tool but not explicitly implemented as \textit{Light Green}. However, we chose against this, as it would trade tighter control and fewer false positives for only a marginal gain in recall. We considered this precision-oriented tradeoff worthwhile, especially since the user always retains the evidence needed to override a conservative call.

Overall, for \textit{Mono2Micro}, the automated mapping proved to be at least as reliable as the manual one, making a few errors of its own but also correcting one that the manual mapping made and flagging uncertainty rather than guessing when expert knowledge was required. Beyond reliability, the skill addresses the fact that manual mapping is time-consuming and error-prone, particularly for tools whose codebase, design, and terminology are unfamiliar to the analyst. Because it can be re-run, cites the evidence behind every finding, and applies the same criteria to every tool, it also makes the mapping more systematic, reproducible, and consistent across analyses, which is essential when evaluating many tools.

Taken together, these results indicate that the skill is reliable and can be used to map other tools onto the feature model.

\subsubsection{Docs-map}
\label{sec:skill-docs-map}

Since Mono2Micro IBM~\cite{Mono2Micro_IBM_2020} is not open source, it was not possible to analyze its codebase. However, online documentation is available, and so the \texttt{docs-map} skill was created to map the features of the model against the information available in this documentation.

This skill mirrors \texttt{codebase-map}, but takes a URL instead of a local path and cites documentation links rather than \texttt{file:line} references as evidence. It produces the same three artifacts.

\subsubsection{Verify-Analysis}
\label{sec:skill-verify-analysis}

Because the mapping skills are non-deterministic and may produce different results, evidence, and conclusions across runs, a third skill, \texttt{verify-analysis}, was created to reconcile them. The intended workflow is to first run \texttt{codebase-map} or \texttt{docs-map} several times on the tool under evaluation and store each mapping in a software configuration management repository, such as Git\footnote{\href{https://github.com/}{https://github.com/}}. The skill then inspects the different versions of a tool's mapping and, whenever discrepancies are found, re-examines the multiple pieces of evidence to determine the most well-supported conclusion. When even this examination is inconclusive, the feature is marked magenta, following the color rules defined in Table~\ref{tab:color-meaning}, and is left for the user to map manually.

This skill takes as input the name of the tool, which must match the name of its analysis, and produces the same three artifacts as the previous skills.

\subsection{Other skills}

\begin{figure}
    \centering
    \includegraphics[width=0.6\linewidth]{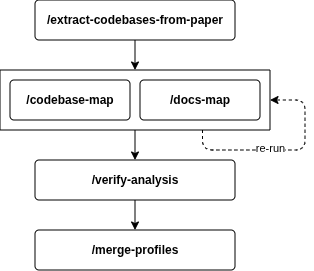}
    \caption{Evaluation\_Pipeline}
    \label{fig:evaluation-pipeline}
\end{figure}

Beyond the skills introduced so far (\textit{Codebase-Map}, \textit{Docs-Map}, and \textit{Verify-Analysis}), two further skills, \textit{Extract-Codebases-From-Paper} and \textit{Merge-Profiles}, were developed to support the analysis of multiple tools. All these skills were integrated into a single evaluation process, illustrated in Figure~\ref{fig:evaluation-pipeline}.

\subsubsection{Extract-Codebases-From-Paper}
\label{sec:skill-extract-codebases-from-paper}

The \texttt{extract-codebases-from-paper} skill makes it easier to obtain the codebases of the tools discussed in a paper. Given a paper, it identifies the tools mentioned in it and, for each one, searches for its codebase or documentation URL. A found codebase is downloaded automatically, while a missing one is recorded so that the user can perform a manual lookup later.

This skill is particularly useful for papers that compare multiple tools, such as~\cite{Wang2024}, as it allows the user to quickly gather the codebases of all the tools discussed. It is also helpful for papers that describe a single tool, since it can cope with broken or outdated links by falling back to an automatic web search.

\subsubsection{Merge-Profiles}
\label{sec:skill-merge-profiles}

Once the mapping skills, \textit{Codebase-Map}, \textit{Docs-Map}, and \textit{Verify-Analysis}, have been run on several tools, one color profile exists per tool. To obtain a broader picture of the feature model's coverage, these per-tool profiles must be combined into a single one. This is the purpose of the \texttt{merge-profiles} skill, which produces a union profile that reflects what a chosen set of tools collectively covers.

The skill can operate in three modes, differing only in which tools it selects. When no arguments are provided, it merges every analyzed tool's profile, capturing the overall coverage achieved across the whole evaluation. Given a paper from which tools were extracted, it merges only that paper's tools, as used to obtain the combined coverage of the tools compared by Wang et al.~\cite{Wang2024}. Finally, given the names of two or more specific tools, it merges only their profiles, which is useful to inspect a particular combination, such as \textit{Mono2Micro} and \textit{Micro2Micro}.

\subsection{Results}
\label{sec:evaluation-results}

Following the methodology described in Section~\ref{sec:evaluation-methodology}, the mapping skills were applied to each of the selected tools mentioned in Table~\ref{tab:evaluation-tools}. Every mapping was then put through an evidence-based review that resolved the inconclusive cases (marked in magenta, as defined in Table~\ref{tab:color-meaning}) and corrected the errors found, producing a revised mapping for each tool. The revised mappings were then merged into several combined profiles.

\begin{table*}[htbp]
\caption{Summary of the mapping results per tool. Each cell reports the number of features assigned that color by the automated mapping and, in parentheses, after the user review.}
\label{tab:mapping-summary}
\centering
\begin{tabularx}{\textwidth}{@{}l X X X X X X @{}}
\toprule
\textbf{Tool} & \textbf{Dark Green} & \textbf{Light Green} & \textbf{Yellow} & \textbf{Red} & \textbf{Orange} & \textbf{Magenta} \\ \midrule
Mono2Micro & 4 (16) & 69 (60) & 24 (23) & 0 (0) & 0 (0) & 2 (0) \\
Micro2Micro & 3 (2) & 44 (44) & 45 (52) & 2 (1) & 0 (0) & 5 (0) \\
Mono2Micro (IBM) & 3 (4) & 52 (44) & 34 (43) & 2 (8) & 0 (0) & 8 (0) \\
HyDec & 1 (1) & 30 (30) & 56 (58) & 11 (10) & 0 (0) & 1 (0) \\
MEM & 3 (3) & 32 (31) & 53 (59) & 2 (1) & 5 (5) & 4 (0) \\
CARGO & 1 (2) & 31 (36) & 48 (54) & 2 (7) & 0 (0) & 17 (0) \\
MOSAIC & 3 (6) & 48 (45) & 45 (47) & 2 (1) & 0 (0) & 1 (0) \\ \midrule
All tools & 15 (17) & 65 (66) & 12 (16) & 0 (0) & 0 (0) & 7 (0) \\\bottomrule
\end{tabularx}
\end{table*}

Table~\ref{tab:mapping-summary} summarizes the mapping results for each tool, reporting, per color, the number of features assigned that color by the automated mapping and, in parentheses, after the user review. At the same time, the final row reports the same counts for the profile obtained by merging all analyzed tools.

The individual mappings and the merged profiles are available in the Appendix~\ref{sec:material} and are discussed in the following sections.

\subsubsection{Mono2Micro}

Starting with \textit{Mono2Micro}~\cite{Lopes2023}, this mapping was very complete, as expected given that it is the tool used for the top-down phase and already discussed in Section~\ref{sec:mapping-validation}. Every feature group was mapped by at least one leaf, including the optional ones such as \textit{Functionality Refactorization} and \textit{Graphical Comparison}, and no constraint was broken.

The tool uses a wide range of granularities, collection sources and techniques, analytical representations, and aggregation criteria, yet implements a single algorithm.

\subsubsection{Micro2Micro}

\textit{Micro2Micro}~\cite{Zhong2025} operates on an existing microservice architecture rather than a monolith and, therefore, maps no monolithic representation at all. This confirms the decision to model the \textit{Representation} of the monolith as optional.

Compared with \textit{Mono2Micro}, \textit{Micro2Micro} covers much less of the feature model, having only a single granularity and a single collection technique, and covers fewer optional features such as \textit{Functionality Refactorization} and \textit{Graphical Comparison}. 
In terms of the mandatory features, it lacks only the \textit{Elements size} metric. This metric is deemed mandatory because it can be extracted from the generated decomposition without any additional computational overhead.

Merging this mapping with the \textit{Mono2Micro} one does not add significant coverage. The \textit{Genetic Algorithm} leaf feature is the only addition, though it confirms the split between mandatory and optional features.

\subsubsection{Wang et al. Tools}

Considering the tools compared by Wang et al.~\cite{Wang2024}, see Table~\ref{tab:evaluation-tools}, each of these tools uses at most two granularities and two collection techniques, and each implements a single algorithm, which denotes a lack of variability in the tools.

\textit{MEM}~\cite{Mazlami2017} breaks a cross-tree constraint by calculating the \textit{Lexical} aggregation criterion from the raw text of source files, without ever holding a representation of the system, either \textit{Structure}, \textit{Call graph}, or \textit{Sequence of accesses}. \textit{MEM} does define an explicit intermediate representation, but its edge weights already encode the aggregation criterion itself. Because the representation and the criterion are collapsed into one artifact, the criterion cannot be recombined with a different one without being recomputed from source and this is precisely the modularity that the constraint is meant to preserve.

Furthermore, none of the five tools uses MoJoFM to compare decompositions, even though it is a mandatory feature of the model. Some tools also lack the \textit{Elements size} metric, namely \textit{CARGO}~\cite{CARGO} and \textit{HyDec}~\cite{Sellami2022}, which makes them break the \textit{Quality Assessment} mandatory constraint. Since one of the main proposed benefits of the model is to support the comparison of tools, this is a significant gap that the model makes explicit and that future tools using it as a reference are positioned to close.

On the other hand, both \textit{MEM}~\cite{Mazlami2017} (with its edge-cutting graph clustering) and \textit{MOSAIC}~\cite{Filippone2023} (with ILP optimization) use algorithms that are not represented in the model. This proves that the model can be easily extended by adding a new leaf with this algorithm, which illustrates its extensibility. The same happens for the \textit{Single Decomposition} subgroup, where \textit{MEM} uses a \textit{Contributor Overlap} metric that is not represented in the model and that can be added as a new leaf.

Once the five tools are merged, coverage is high, with only the optional \textit{Functionality Refactorization} group left unmapped. The merged profile shows how the feature model can accommodate a diverse set of tools that had already been compared with one another, presenting that comparison in a more systematic and visual way.

\subsubsection{Overall Analysis}

The research question in Section~\ref{sec:research-questions} was extended to also include microservices refactorization, as explained in Section~\ref{sec:representation} and is now "How can we identify and represent the common parts and variability of the approaches for the identification and refactoring of microservices in monolithic and microservice systems?"

To answer this research question, we merged the profiles of all the analyzed tools to have a complete picture of all the existing features implemented by the reference tools.
The merged profile, available in the Appendix~\ref{sec:material} and reported in the last row of Table~\ref{tab:mapping-summary}, yields a version of the feature model in which the great majority of features are mapped and no mandatory feature is left unmapped.

Most of the sources are covered, and the few that are not, together with the unmapped \textit{Model Analysis} collection technique, should not be read as features that do not belong in the model. On the contrary, they reveal a gap in the existing tools, since these same sources and collection techniques are discussed in the literature, as our analysis of the surveys and systematic reviews shows.

The same happens for algorithms: \textit{Genetic algorithms}, \textit{Community detection}, and \textit{Hierarchical clustering} are the most frequently implemented, while \textit{K-means++}, \textit{Formal concept analysis}, and the \textit{Hungarian algorithm} appear in none of the analyzed tools, whereas \textit{MEM}~\cite{Mazlami2017} contributes one algorithm outside the model. This makes the pattern from the individual tools explicit: implementing some algorithm is a common part shared by every tool, whereas the choice of a specific algorithm is a variation point. Since all the \textit{Quality Assessment} features are realizable, a tool that implements several algorithms will allow their systematic comparison.

Comparing the numbers of each tool in Table~\ref{tab:mapping-summary} with those of the merged profile shows that the tools are complementary: they cover different parts of the model, so the merged profile reaches more mapped features than any individual tool while leaving no mandatory feature unmapped. This confirms that the model can accommodate a diverse set of already implemented approaches, allowing their comparison in a systematic and visual way.

Taken together, this fully merged profile answers the research question. The mandatory feature groups capture the common parts shared by the tools, while the optional leaves capture the variability points, which can themselves be extended with new leaves, as happened with the algorithms discovered during the analysis.

\section{Related Work}
\label{sec:related-work}

To the best of our knowledge, this is the first analytical description proposed for the problem space of microservice identification. Consequently, our related work analysis is limited to existing mapping studies and systematic literature reviews. 

These studies exhibit distinct focuses:~\cite{Capuano2022} examines the quality attributes of the approaches;~\cite{Abdellatif2021} provides a detailed breakdown of the elements comprising different approaches;~\cite{Trabelsi2025} investigates the problem from a machine learning perspective;~\cite{Abgaz2023} categorizes its analysis by migration phases;~\cite{Oumoussa2024} reviews the state of the art, concluding that comparing techniques remains a key challenge;~\cite{Auer2021} emphasizes quality aspects, particularly metrics;~\cite{Yang2025} addresses a broad range of factors in the microservices lifecycle, where decomposition is just one component;~\cite{Saucedo2025} dissects the migration process, concluding that no single technique is entirely effective;~\cite{Mohottige2025} provides an in-depth review highlighting the need for further work on evaluation;~\cite{Nassima2025} focuses on semantic approaches, such as NLP, ontologies, and AI models; and~\cite{Oumoussa2025} specifically targets ontology-based methods.

Although these prior studies do not provide a formal synthesis, we can still analyze whether and how each addresses the core features of our proposed model.

The literature shows that different approaches operate at various levels of abstraction, ranging from fine-grained elements, such as methods~\cite{Abdellatif2021} and classes~\cite{Abgaz2023, Mohottige2025, Trabelsi2025}, to coarser-grained elements, such as entities~\cite{Abgaz2023, Abdellatif2021}, functionalities~\cite{Mohottige2025}, packages~\cite{Trabelsi2025}, or API endpoints~\cite{Saucedo2025, Yang2025}. However, while existing studies acknowledge the relevance of abstraction levels, they fail to recognize granularity as a variation point that impacts the characteristics of an approach. Consequently, they overlook how crucial this variation is when comparing different approaches.

In what concerns the \textit{Representation Collection} features, the studies consider artifacts, such as source code~\cite{Abgaz2023, Abdellatif2021, Mohottige2025, Capuano2022, Trabelsi2025, Saucedo2025}, execution traces~\cite{Abgaz2023, Abdellatif2021, Mohottige2025, Yang2025, Capuano2022, Trabelsi2025, Saucedo2025}, business models~\cite{Abdellatif2021, Mohottige2025, Saucedo2025, Yang2025, Capuano2022, Trabelsi2025}, or digital documentation~\cite{Abdellatif2021, Abgaz2023, Saucedo2025, Trabelsi2025}. Moreover, they identify \textit{Collection Techniques} like \textit{Static Analysis}~\cite{Abgaz2023, Abdellatif2021, Capuano2022, Oumoussa2024, Nassima2025, Oumoussa2025, Trabelsi2025}, \textit{Dynamic Analysis}~\cite{Abdellatif2021, Mohottige2025, Yang2025},  \textit{Version Analysis}~\cite{Abdellatif2021, Mohottige2025},  and \textit{Model Analysis}~\cite{Abdellatif2021, Mohottige2025, Saucedo2025, Trabelsi2025, Yang2025}. However, they lack a clear separation between the \textit{Source} and the \textit{Collection Technique} hiding the possible combinations. Therefore, the secondary literature provides the raw classifications of what artifacts are collected, but they remain primarily descriptive. They lack a multi-layered subfeature hierarchy that allows a software architect to see the explicit constraints between features, such as how a specific \textit{Collection Technique} strictly requires a specific type of source (e.g., \textit{Dynamic Analysis} requiring a \textit{Runtime Source}).

The same problem occurs in the \textit{Analytical} representation features. The studies identify their elements, such as the internal composition of elements~\cite{Trabelsi2025, Mohottige2025}, as well as inter-structure relationships, including inheritance~\cite{Trabelsi2025, Mohottige2025} and association dependencies~\cite{Trabelsi2025, Yang2025, Abgaz2023}. They also identify representations of runtime interactions between system elements, such as call graphs~\cite{Abgaz2023, Trabelsi2025, Yang2025}, sequences of accesses~\cite{Trabelsi2025}, and resource-related information~\cite{Abgaz2023}, such as CPU usage or response time. Moreover, some studies identify \textit{Evolution} representations~\cite{Abgaz2023, Mohottige2025}. However, there is no explicit separation of representation, which hinders the modularity, understandability, and comparability of the approaches.

Furthermore, and perhaps more importantly, existing literature focuses exclusively on identifying microservices within monolithic systems, largely ignoring the refactoring of existing microservice architectures. To address this gap, our feature model fosters reusability because both activities are tightly interconnected in terms of their objectives and techniques.

Another consequence of addressing both monolith decomposition and microservice refactoring is recognizing that visualizations are largely ignored in existing surveys. For instance, the work of Nakazawa~\cite{Nakazawa2018} is omitted. By introducing a \textit{Visualization} feature, our model enables the definition of explicit cross-tree dependencies. For example, it dictates that an architect cannot manually perform a merge of clusters action unless the visualization tool specifically supports a structural cluster view.

The literature considers several \textit{Aggregation Criteria}, such as \textit{Task} and \textit{Contributor} criteria~\cite{Abgaz2023}, \textit{Lexical}~\cite{Nassima2025, Oumoussa2025, Yang2025}, \textit{Functional}~\cite{Abgaz2023, Oumoussa2024, Oumoussa2025, Yang2025}, and \textit{Adjacency}~\cite{Abgaz2023, Oumoussa2024, Oumoussa2025}. However, only a few studies consider \textit{Expert knowledge}~\cite{Abgaz2023, Abdellatif2021, Saucedo2025, Trabelsi2025, Yang2025}.

The \textit{Algorithms} feature is also extensively described in the literature: \textit{Hierarchical Clustering}~\cite{Abgaz2023, Abdellatif2021, Trabelsi2025}, \textit{Community Detection}~\cite{Oumoussa2024, Nassima2025, Oumoussa2025}, \textit{K-means++}~\cite{Abdellatif2021, Yang2025}, \textit{Formal Concept Analysis}~\cite{Abgaz2023, Saucedo2025}, \textit{Genetic Algorithms}~\cite{Yang2025, Trabelsi2025}, and the \textit{Hungarian Algorithm}~\cite{Abgaz2023}. However, \textit{Manual Editing} is not addressed, as the lack of the \textit{Visualization} feature, just mentioned above.

The analyzed literature effectively treats metrics as a separate aspect. Metrics such as \textit{Cohesion}~\cite{Abgaz2023, Abdellatif2021, Mohottige2025, Saucedo2025, Capuano2022, Oumoussa2024,Trabelsi2025}, \textit{Coupling}~\cite{Abgaz2023, Abdellatif2021, Mohottige2025, Saucedo2025, Capuano2022, Oumoussa2024, Trabelsi2025}, \textit{Complexity}~\cite{Saucedo2025, Trabelsi2025}, and \textit{Team Size}~\cite{Mohottige2025, Saucedo2025}. 

Only a select few studies attempt to extensively categorize existing methodologies~\cite{Abdellatif2021, Abgaz2023, Trabelsi2025, Saucedo2025, Mohottige2025}, but these efforts lack a comprehensive analytical framework to benchmark current approaches and guide future advancements.

Overall, the primary drawback of existing surveys and systematic literature reviews is that they are fundamentally descriptive rather than analytical. Because they merely catalog and describe standalone monolith-to-microservice identification approaches, they do not establish a common conceptual framework. Consequently, researchers and practitioners must perform significant manual preprocessing to normalize terminology and results across different techniques just to understand how they relate to one another. This lack of a unified analysis framework is precisely why the field has seen such a high volume of secondary studies published in under a decade without achieving a clear, unified direction.

Furthermore, surveys fail to provide actionable, practical tooling to support experimentation, comparison, or the integration of different architectural approaches. By treating existing tools as rigid, standalone solutions, surveys highlight approaches that are often locked into a single specific level of granularity - such as only analyzing classes or only analyzing APIs - with no mechanism to combine strategies across different conceptual levels. Because they lack this capability, surveys offer no concrete roadmap for software architects who need to perform multi-objective trade-off analyses (e.g., balancing performance against modularity), a task that inherently requires integrating multiple, disparate approaches to find an optimal system decomposition.

\section{Threats to Validity}
\label{sec:threats-to-validity}

\subsection{Construct Validity}

Because terminology differs across the analyzed papers, the construction of the feature model required interpreting and consolidating concepts that looked distinct on the surface but were semantically close. This introduces subjectivity, but we reduced it by following the principles of Nešić et al.~\cite{Nesic2019}, combining a top-down phase with an iterative bottom-up phase and documenting the resulting features in this paper.

Making \textit{Metrics} subgroup mandatory as a child of \textit{Quality Assessment} further assumes that decomposition quality is something a tool can compute. However, as stated in multiple papers~\cite{Wang2024, Oumoussa2024, Nassima2025}, solely relying on metrics to assess the quality of a decomposition is insufficient. These metrics are a proxy for quality, and the problem is compounded by different tools computing the same metric over different information, as Table~\ref{tab:metrics-dependencies} makes explicit. An expert architectural review is still required to judge whether service boundaries are sensible.

\subsection{Internal Validity}

The codebase-map (Section~\ref{sec:skill-codebase-map}) and docs-map (Section~\ref{sec:skill-docs-map}) skills are non-deterministic, so their output has to be validated before we rely on it. Every finding must be backed by explicit evidence, and we later reconciled the findings across multiple runs with the verify-analysis skill (Section~\ref{sec:skill-verify-analysis}). Even then, a domain expert validated the resulting mapping. That expert check matters even more given that we took the deliberate design decision of not assigning the Light Green color, Table~\ref{tab:color-meaning}, when a feature is present in principle but not explicitly implemented by the tool.

\subsection{External Validity}

The generalizability of the proposed feature model depends on the sources used for its design. The evaluation showed that new features can be incorporated as needed. Consequently, tools and studies outside the selected scope may expose missing features, some of which might not fit the current model easily. Furthermore, the automated mapping of documentation (docs-map) depends heavily on the quality of the available documentation. On the other hand, the quality of the selected surveys ensures that the feature model is built upon the state of the art. Additionally, our ability to homogeneously analyze reference tools further validates the model's quality.

\subsection{Conclusion Validity}

We claim that the automated mapping is at least as reliable as a manual one, as long as an expert verifies it. In the Mono2Micro validation, the automated mapping caught a human error that the manual mapping had made. This claim rests on a single tool for which a reference manual mapping was available, so confirming it more broadly would require testing across more tools.

\section{Conclusion}
\label{sec:conclusion}

As organizations increasingly migrate from monolithic architectures to microservices to achieve better scalability, maintainability, and team autonomy, they face the fundamental challenge of correctly identifying service boundaries. While the past decade has seen a rapid proliferation of microservice identification approaches, this expansion has resulted in a highly fragmented research landscape. Different methodologies rely on distinct data collection techniques, operate at varying levels of conceptual granularity, and prioritize different quality attributes.

Consequently, the primary challenge in the current research is no longer a lack of decomposition algorithms, but rather the absence of a unified analytical framework to support them. Existing secondary studies remain primarily descriptive, and current identification tools are built as rigid, standalone solutions. This fragmentation severely limits systematic comparison, hinders continuous experimentation, and prevents software architects from performing the multi-objective trade-off analyses necessary to balance competing architectural requirements.

The primary advantage of the feature model is that it shifts the field from passive observation to active engineering by providing a unified analysis framework. While traditional surveys simply describe and categorize existing solutions, the feature model explicitly maps out the entire "problem space" of microservice decomposition. This comprehensive mapping acts as a practical blueprint for designing variant-rich identification tools, allowing developers to build modular systems where different data collection methods, clustering algorithms, and quality metrics can be seamlessly swapped and tested against one another. By making this variability explicit, the model creates a roadmap for benchmarking that highlights unexplored research areas and standardizes how future strategies are compared.

Furthermore, the feature model is uniquely equipped to handle the messy reality of real-world legacy systems and complex architectural decisions. It explicitly decouples the initial data sources from the resulting analytical representations. This means that if a legacy system is missing a specific architectural source, such as up-to-date documentation, the model can flexibly generate the same representation using an alternative source, like runtime logs, adapting to what is actually available. Additionally, the model facilitates multi-objective trade-off analysis, enabling software architects to integrate disparate techniques, like combining business-process mining with static code analysis, to successfully balance conflicting system requirements such as performance and modularity.

As future work, we plan to implement the proposed feature model as an executable software product line, itself implemented as a microservices system, thereby extending its current role beyond serving as an analysis and comparison framework. In a first phase, this implementation will target monolith-to-microservices identification, but the same framework is intended to generalize to microservices-to-microservices identification, where a previously proposed microservice architecture is re-decomposed. Implementing the model as a software product line will enable the validation of its completeness and practical applicability while supporting the rapid prototyping and empirical evaluation of new microservice identification techniques. By realizing the modeled variability as composable implementation artifacts, each deployable as an independent service, the framework could support the derivation of identification tools through feature-based configuration, enabling alternative data collection methods, representations, decomposition algorithms, and quality metrics to be systematically combined and empirically compared within a common architectural framework.

\section{Material}
\label{sec:material}

All the material supporting this study is available as a browsable appendix in ~\cite{UnifiedFeatureModelArtifacts}. It consists of the following artifacts:

\begin{itemize}
  \item \textbf{Unified feature model}, the final model presented in this paper, in the machine-readable FeatureIDE format together with its rendered image, so that it can be inspected, configured, and extended.
  \item \textbf{Initial feature model}, the first draft obtained in the top-down phase of the modeling process (Section~\ref{sec:modeling-process}) from the analysis of \textit{Mono2Micro}~\cite{Lopes2023}, before the generalization introduced by the bottom-up phase.
  \item \textbf{Meta-review data}, the meta-review table over the 11 selected secondary studies, in both spreadsheet and document form.
  \item \textbf{Mapping skills}, the Claude skills used to produce the mappings, namely \texttt{codebase-map} (Section~\ref{sec:skill-codebase-map}), \texttt{docs-map} (Section~\ref{sec:skill-docs-map}), and \texttt{verify-analysis} (Section~\ref{sec:skill-verify-analysis}), together with the auxiliary \texttt{extract-codebases-from-paper} (Section~\ref{sec:skill-extract-codebases-from-paper}) and \texttt{merge-profiles} (Section~\ref{sec:skill-merge-profiles}) skills.
  \item \textbf{Evaluation data}, for each of the seven tools mapped in the evaluation (Table~\ref{tab:evaluation-tools}), the analysis and the color profile generated by the mapping skills, and the revised profile resulting from the user review of that mapping, as well as the merged profiles. It also includes the exported images of every mapping and, for the tools whose mapping extended the model, the resulting extended FeatureIDE models.

\end{itemize}

\section*{Acknowledgment}

\noindent
\small The authors acknowledge the use of Claude and Gemini to improve the English phrasing and readability of this manuscript. Additionally, Claude skills were utilized for the evaluation methodology as described in Section~\ref{sec:evaluation-methodology}.

\noindent 
\small Work supported by national funds through Fundação para a Ciência e a Tecnologia, I.P. (FCT) under projects UID/50021/2025 (DOI: \href{https://doi.org/10.54499/UID/50021/2025}{https://doi.org/10.54499/UID/50021/2025}) and UID/PRR/50021/2025 (DOI: \href{https://doi.org/10.54499/UID/PRR/50021/2025}{https://doi.org/10.54499/UID/PRR/50021/2025}).

\bibliographystyle{unsrt}
\bibliography{mybib}


\end{document}